\documentclass[10pt]{article}

\pdfoutput=1 

\usepackage{jheppub} 

\usepackage[T1]{fontenc} 

\usepackage{amssymb,amsthm,cancel,hyperref,graphicx,xcolor}
\usepackage{mathrsfs,wasysym}
\usepackage{booktabs}
\usepackage{tikz}
\usetikzlibrary{calc}
\usepackage{placeins}

\newcommand{\sss}{\scriptscriptstyle}

\newcommand{\as}{\alpha_s}

\newcommand{\Ord}{\mathcal{O}}

\newcommand{\muf}{\mu_{\sss\rm F}}
\newcommand{\mur}{\mu_{\sss\rm R}}

\newcommand{\gammae}{\gamma_{\scriptscriptstyle E}}

\let\originalleft\left
\let\originalright\right
\renewcommand{\left}{\mathopen{}\mathclose\bgroup\originalleft}
\renewcommand{\right}{\aftergroup\egroup\originalright}
\newcommand{\DYqt}{\texttt{DYqT}}

\newcommand{\Gt}{\mathcal{G}_\text{thr}}
\newcommand{\Gj}{\mathcal{G}_\text{joint}}
\newcommand{\Gq}{\mathcal{G}_{Q_T}}

\def\beq{\begin{equation}}  
\def\eeq{\end{equation}}
\def\({\left(}
\def\){\right)}
\def\[{\left[}
\def\]{\right]}

\title{\boldmath Vector boson production in joint resummation}

\author[a]{Simone~Marzani,}
\author[a]{Vincent~Theeuwes.}

\affiliation[a]{University at Buffalo, The State University of New York, Buffalo, NY 14260-1500, USA}

\emailAdd{smarzani@buffalo.edu}
\emailAdd{vtheeuwe@buffalo.edu}

\abstract{We study the transverse momentum ($Q_T$) distribution of an electro-weak vector boson produced via the Drell-Yan mechanism, in the context of joint resummation. This formalism allows for the simultaneous resummation of logarithmic contributions that are enhanced at small $Q_T$ and at partonic threshold. We extend joint resummation to next-to-next-to leading logarithmic accuracy and we present resummed and matched results for three different phenomenological setups. In particular, we study the production of a $Z$ boson at the Tevatron and at the Large Hadron Collider (LHC), as well as the production of a heavier $Z^\prime$ at the LHC. We compare our findings to standard $Q_T$ resummation, as well as to fixed-order perturbation theory. We find that joint resummation provides a moderate (but not flat) correction with respect to $Q_T$ resummation and it leads to a reduction of the scale dependence of the results. However, our study also shows some limitations of this formalism. While the use of joint resummation for $Z$ production at the Tevatron and $Z^\prime$ production at the LHC appears to be justified, our implementation suffers from a stronger dependence on power corrections for processes which are further away from threshold, such as $Z$ production at the LHC, for which we cannot claim an improvement over standard $Q_T$ resummation.
}

\begin{document}
\maketitle

\section{Introduction}\label{sec:intro}
Last year the CERN Large Hadron Collider (LHC) started its second run of operation colliding protons at 13 TeV. 
While convincing hints of new physics are still eluding the experiments,  one of the main goals of Run II remains precision physics near the electro-weak scale. These analyses are performed with an increasing amount of data, making accurate theoretical predictions for differential distributions more relevant than ever. 
One of the most extensively studied distributions at hadron colliders is the transverse momentum ($Q_T$) spectrum of electro-weak bosons produced via the Drell-Yan (DY) mechanism~\cite{Affolder:1999jh,Abbott:1999yd,Abazov:2007ac,Abazov:2008ez,Abazov:2010kn,Abazov:2010mk,Aad:2011gj,Chatrchyan:2011wt,Aaij:2012mda,Aad:2014xaa,Aaij:2015vua,Khachatryan:2015oaa,Aaij:2015gna,Aad:2015auj,Khachatryan:2016nbe}. 
Studies of $Q_T$ spectra and related angular correlations of DY lepton pairs provide a useful testing ground for an even more interesting Higgs and new physics program.
Remarkably, the accuracy of LHC measurements in the context of electro-weak boson distributions has now reached the percent level~\cite{Aaij:2015gna,Aad:2015auj,Khachatryan:2016nbe}. Consequently, substantial effort has been recently put by the theory community in order to shrink the uncertainty affecting theoretical predictions. High-precision fixed-order calculations, which have been recently performed to next-to-next-to-leading order (NNLO) accuracy~\cite{Boughezal:2015aha,Boughezal:2015dra,Boughezal:2015dva,Boughezal:2013uia,Chen:2014gva,Ridder:2015dxa}, can be employed to describe the moderate-to-large region of the $Q_T$ spectrum.
However, the small transverse momentum region is dominated by the emission of soft and collinear partons, and it is characterized by the presence of large logarithms of $Q_T/Q$, where $Q$ is the invariant mass of the final state, which need to be resummed. Thus, reliable predictions across a vast range of $Q_T$ values can be obtained by matching fixed-order and resummed predictions.

While providing an all-order prediction for the shape of transverse momentum spectrum, $Q_T$ resummation carries very little information about its normalization, beyond the fixed-order that it is matched to.
On the other hand, the determination of inclusive cross sections can be improved beyond fixed order by including threshold resummation.
Partonic coefficient functions contain plus distributions which exhibit logarithmic enhancement in the variable $z=1-Q^2/\hat{s}$ where $\hat{s}=x_1 x_2 s$ is the partonic center of mass energy squared. Even though the collision energy of the protons is much larger than the electro-weak scale, these contributions can still be large because parton distribution functions (PDFs) at large $Q^2$ preferentially sample the region of low momentum fractions $x_1$ and $x_2$.

Transverse momentum and threshold logarithms originate from the emission of soft gluons. Therefore, it is natural to look for a framework that allows for a consistent resummation of both. 
The general formalism to perform this joint resummation was derived some time ago~\cite{Li:1998is,Laenen:2000ij} and explicitly worked out to next-to-leading logarithmic (NLL) accuracy in both variables. It was then applied to a number of phenomenological studies including prompt photon~\cite{Laenen:2000ij}, Drell-Yan~\cite{Kulesza:2002rh}, Higgs~\cite{Kulesza:2003wn}, top pair~\cite{Banfi:2004xa} and electro-weak supersymmetric particle~\cite{Bozzi:2007tea,Fuks:2013vua} production. 
Moreover, the universality properties of transverse momentum and threshold resummation have been extensively discussed in Ref.~\cite{Catani:2013tia}.
However, despite the fact that both threshold and $Q_T$ resummation are known to at least NNLL, the application of this method has not yet been extended to higher logarithmic accuracy.
We also note that recent progress has also been made in describing joint resummation in the context of the Soft-Collinear-Effective Theory (SCET)~\cite{Li:2016axz,Lustermans:2016nvk}. In addition, threshold resummation has been included into a parton shower, which also allows the description of the small transverse momentum region~\cite{Nagy:2016pwq}.

In this paper we concentrate on the case of vector boson production via the DY mechanism and we extend the study of Ref.~\cite{Kulesza:2002rh} to NNLL accuracy. 
This paper is organized as follows.
We begin in Section~\ref{sec:recap} with a brief overview of threshold and $Q_T$ resummations. We then describe joint resummation in Section~\ref{sec:combination}, first reviewing the NLL case and then extending it to NNLL accuracy. Next, in Section~\ref{sec:numerics} we present our numerical results together with a study of the reliability of the approximations employed in our implementation of joint resummation, before concluding in Section~\ref{sec:conclusion}. More technical details are collected in the Appendices. 

\section{A recap of transverse momentum and threshold resummations}\label{sec:recap}
In this section we provide a brief overview of threshold and transverse-momentum resummations.
In addition, the all-order results will be written in a way that allows for a comparison between the two which eases their generalization to joint resummation.

\subsection{Threshold resummation}\label{sec:threshold}
Threshold resummation was originally introduced at the end of the 1980s~\cite{Sterman:1986aj,Catani:1989ne}. After that it was extended to NNLL accuracy~\cite{Vogt:2000ci,Catani:2003zt}, and even to N$^3$LL accuracy, e.g.~\cite{Moch:2005ba, Moch:2005ky, Laenen:2005uz, Bonvini:2014joa,Catani:2014uta,Bonvini:2014tea,Schmidt:2015cea,Bonvini:2016frm} for electro-weak final states.  Moreover, the resummation of large threshold logarithms has also been formulated using SCET, see e.g.~\cite{Becher:2007ty}.

In this study we concentrate on the production of a  (neutral) vector boson $F$ and we are interested in resumming logarithms of $1-Q^2/\hat{s}$, where $\sqrt{\hat{s}}$ is the partonic center-of-mass energy and $Q$ is the vector boson invariant mass. 
Threshold resummation is usually performed in Mellin space, where the threshold limit corresponds to $N \to \infty$ and large logarithms of $1-Q^2/\hat{s}$ are mapped into logarithms of $N$.
The cross section can be written as
\begin{align}
\sigma_F(s,Q^2)&=\sum_{a}\sigma^{(0)}_{a\bar{a}\to F}\int_{C_{T}}\frac{dN}{2\pi i} \left(\frac{Q^2}{s}\right)^{-N+1}\tilde{f}_{a/h_1}(N,\muf^2)\;\tilde{f}_{\bar{a}/h_2}(N,\muf^2)\nonumber\\
&\times G_{a\bar{a}}(N,\as(\mur),Q^2/\mur^2,Q^2/\muf^2),
\end{align}
where $\sigma_{a\bar{a}\to F}^{(0)}$ is the lowest order cross section for the partonic process $a\bar{a}\to F$ and the parton densities are indicated by $f_{a/h}(x,\mu^2)$. We have also introduced the renormalization scale $\mur$ and the factorization scale $\muf$, while $C_{T}$ indicates the contour for the inverse Mellin transform. At leading power, threshold resummation does not receive any contribution from initial-state off-diagonal flavor components, therefore the only partonic subprocess that contributes is $a\bar{a}$.  The function $G_{a\bar{a}}$ is given by
\begin{equation}
G_{a\bar{a}}(N,\as(\mur),Q^2/\mur^2,Q^2/\muf^2)=\mathcal{C}_{a\bar{a}}(\as(\mur),Q^2/\mur^2,Q^2/\muf^2)\;\exp\left[\Gt(N,\as(\mur),Q^2/\mur^2,Q^2/\muf^2)\right],
\end{equation}
where we have introduced the Sudakov exponent~\cite{Catani:2003zt}
\begin{align}\label{eq:sudakov-thr}
\Gt(N,\as(\mur),Q^2/\mur^2,Q^2/\muf^2)=&-\int_{N_0/N}^{1}\frac{dy}{y}\left[2\int^{y^2 Q^2}_{\mu_F^2}\frac{dq^2}{q^2}A_{a}(\as(q))+\tilde{D}_{a}(\as(yQ))\right]\nonumber \\
=&\int^{Q^2}_{Q^2/\bar{N}^2}\frac{dq^2}{q^2}\left[2A_{a}(\as(q))\log\left(\frac{\bar{N}q}{Q}\right)-\frac{1}{2}\tilde{D}_{a}(\as(q))\right] \nonumber \\
& -2\log\bar{N}\int_{\muf^2}^{Q^2}\frac{dq^2}{q^2}A_{a}(\as(q)),
\end{align}
with $\bar N=N/N_0=Ne^{\gammae}$. The functions $A$,  $\tilde{D}$ and $\mathcal{C}$ admit a perturbative expansion in the strong coupling
\begin{align}
A_a(\as) = \sum_{n=1}^{\infty} \left(\frac{\as}{\pi}\right)^n A_a^{(n)}\,,\qquad&
\tilde{D}_a(\as) = \sum_{n=2}^{\infty} \left(\frac{\as}{\pi}\right)^n \tilde{D}_a^{(n)},
\end{align}
\begin{align}\label{CfuncTH}
\mathcal{C}_{a\bar{a}}(\as) = 1+\sum_{n=1}^{\infty} \left(\frac{\as}{\pi}\right)^n \mathcal{C}_{a\bar{a}}^{(n)}\,.
\end{align}
The function $A_a$ is the cusp anomalous dimension, $\tilde{D}_a$ accounts for soft emissions at large angle, while $\mathcal{C}_{a\bar{a}}$ takes into account the virtual corrections. Explicit expressions are collected in Appendix~\ref{sec:coef}.
The perturbative order at which the above coefficients are included determines the logarithmic accuracy of the result. Throughout this paper we adopt a logarithmic counting in the exponent $\mathcal{G}$. Therefore, N$^k$LL accuracy is achieved if $A_a$ is included up to (and included) $\Ord \left(\as^{k+1} \right)$, $\tilde{D}_a$ up to $\Ord \left(\as^{k} \right)$ and $\mathcal{C}_a$ up to $\Ord \left(\as^{k-1} \right)$. Moreover, the accuracy can be promoted to N$^k$LL$^\prime$ if also the  $\Ord \left(\as^{k} \right)$ contribution to $\mathcal{C}_a$ is included. We keep the same convention for $Q_T$ and joint resummation.

While in this paper we concentrate on NNLL accuracy, these coefficients have actually been computed to high-enough accuracy to achieve N$^3$LL$^\prime$ accuracy, with the exception of the four-loop contribution to the cusp. They can be found in \cite{Catani:1989ne,Catani:1990rr,Catani:1998tm,Vogt:2000ci,Catani:2001ic,Catani:2003zt,Moch:2004pa,Vogt:2004mw,Moch:2005ba}. Moreover, in order to achieve the desired logarithmic accuracy, the integrals over the QCD running coupling $\as(q)$ must be performed with the $\beta$ function at the appropriate perturbative order.

\subsection{$Q_T$ resummation} \label{sec:qt}

Since the original paper on $Q_T$ resummation \cite{Collins:1984kg} a lot of effort has gone into further improving the accuracy of theoretical predictions in order to perform meaningful comparisons to experimental results. Resummed results at NNLL$^\prime$ matched to NLO have been available for quite some time, see e.g.~\cite{Bozzi:2003jy,Bozzi:2005wk,Bozzi:2007pn,Catani:2010pd,Gehrmann:2014yya,Monni:2016ktx}. The resummation of small $Q_T$ logarithms has also been formulated in the context of SCET~\cite{Gao:2005iu,Idilbi:2005er,Mantry:2009qz,Mantry:2010mk,Becher:2010tm,GarciaEchevarria:2011rb,Becher:2012yn,Chiu:2012ir,Neill:2015roa,Ebert:2016gcn}.
In addition, several computer codes have been developed for $Q_T$ resummation at this accuracy in the case of neutral boson production, e.g.~\cite{Ladinsky:1993zn,Landry:2002ix,Bozzi:2008bb,Bozzi:2010xn,Banfi:2012du,deFlorian:2011xf,deFlorian:2012mx}.
Recently, the calculation of the NNLO corrections to the $Q_T$ distribution of neutral boson production processes has been completed~\cite{Boughezal:2015aha,Boughezal:2015dra,Boughezal:2015dva,Boughezal:2013uia,Chen:2014gva,Ridder:2015dxa}, and N$^3$LL precision is within reach~\cite{Li:2016ctv}.

$Q_T$ resummation is usually performed in Fourier space with $b$ being the variable conjugate to the transverse momentum $Q_T$. In this conjugate space the small $Q_T$ limit corresponds to the large $b$ limit. The resummed transverse momentum distribution of an electroweak final state $F$ can be written as
\begin{equation}
\frac{d\sigma_{F}}{dQ_T^2} = \frac{Q^2}{s}\int_{0}^{\infty}db\frac{b}{2}J_0(bQ_T)W^F(b,Q,s).
\end{equation}
The goal of this study is to understand the similarities in the structure of $Q_T$ and threshold resummation in order to combine these two methods of resummation. It is easier to understand the overlap and differences between the two if we transform the expression for $Q_T$ resummation into Mellin space with respect to $z=Q^2/s$:
\begin{align}
\tilde{W}^F(N,b,Q) &= \int_{0}^{1}dz \; z^{N-1}\;W^F(b,Q,s)\nonumber\\
&=\; \sum_{a}\sigma_{a\bar{a}\to F}^{(0)}(\as(Q))H^F_a(\as(Q))S_a(Q,b)\nonumber\\
&\times \; \sum_{c,d}\tilde{C}_{ac}(N,\as(Q/{\bar b}))\tilde{C}_{\bar{a}d}(N,\as(Q/{\bar b}))\tilde{f}_{c/h_1}(N,Q^2/{\bar{b}}^2)\tilde{f}_{d/h_2}(N,Q^2/{\bar{b}}^2)
\end{align}
where $\bar b= Q b/b_0$, with $b_0=2 e^{-\gammae}$. 
The function $S_a(Q,b)$ is the Sudakov form factor for particle $a$ and is given by:\footnote{Note that a resummation scale $\mu_Q$ is often introduced in the context of $Q_T$ resummation as a means to estimated the size of higher-order logarithmic corrections. On the other hand, this kind of variation is not usually considered for threshold resummation. Therefore, here we fix $\mu_Q=Q$, while we still allow for renormalization ($\mur$) and factorization ($\muf$) scale variations.}.
\begin{equation}
S_a(Q,b)=\exp\left\{-\int_{Q^2/{\bar b}^2}^{Q^2}\frac{dq^2}{q^2}\left[A_a(\as(q))\log\frac{Q^2}{q^2}+B_a(\as(q))\right]\right\}
\end{equation}
The functions $A$, $B$, $C$ and $H^F$ are given as perturbative series in $\as$ and can be found in \cite{Kodaira:1981nh,Kodaira:1982az,Davies:1984hs,Vogt:2000ci,Berger:2002sv,Moch:2004pa,Vogt:2004mw,Catani:2009sm,Becher:2010tm} (see also Appendix~\ref{sec:coef}):
\begin{align}\label{QTcoeff}
A_a(\as) = \sum_{n=1}^{\infty} \left(\frac{\as}{\pi}\right)^n A_a^{(n)}\,,\qquad&
B_a(\as) = \sum_{n=1}^{\infty} \left(\frac{\as}{\pi}\right)^n B_a^{(n)}\, , \nonumber \\
H^F_a(\as) = 1+\sum_{n=1}^{\infty} \left(\frac{\as}{\pi}\right)^n H_a^{F,\;(n)}\,,\qquad&
\tilde{C}_{ac}(N,\as) = \delta_{ac}+\sum_{n=1}^{\infty} \left(\frac{\as}{\pi}\right)^n \tilde{C}_{ac}^{(n)}(N)\,
\end{align}
The coefficients $A_a^{(1)}$ and $A_a^{(2)}$ are the same as for threshold resummation and they correspond to the cusp anomalous dimension. However, starting from $A_a^{(3)}$ this contribution becomes observable-dependent, see. e.g~\cite{Becher:2010tm,Monni:2011gb}, and, in particular, it is different for $Q_T$ and threshold resummations. Henceforth, we will denote these coefficients with $A_{a,\;Q_T}^{(3)}$ and $A_{a,\;\mathrm{thr}}^{(3)}$, respectively.

Furthermore, in order to facilitate the comparison to threshold resummation we run the PDFs to the factorization scale $\mu_F$, we evaluate the coefficient functions $C_{ij}$ at the hard scale of the process $Q$ and we combine the hard functions, $H^F$, $C$ and $\sigma^{(0)}$, into one perturbative hard factor. 
Because of the non-diagonal nature in flavor space of the Dokshitzer-Gribov-Lipatov-Altarelli-Parisi (DGLAP) splitting functions and of the closely related $C_{ab}$ functions, $Q_T$ resummation requires dealing with path-ordered exponentials of the anomalous dimension matrix. 
In order to simplify our discussion, we consider here only one flavor-diagonal contribution, which is the one that is enhanced at threshold and we drop all flavor indices.
With this simplification, we obtain
\begin{align} \label{WqT}
\tilde{W}^F(N,b,Q) &=\; \mathcal{H}^{F}(N,Q,\as(\mur),Q^2/\mur^2,Q^2/\muf^2) \;\tilde{f}_{a/h_1}(N,\muf^2)\;\tilde{f}_{\bar a/h_2}(N,\muf^2)\nonumber\\
&\times \;\exp\left\{\Gq(N,\as(\mur),b,Q^2/\mur^2)\right\}.
\end{align}
where the hard-function is now defined by
\begin{align}\label{HfuncQT}
&\mathcal{H}^{F}(N,Q,\as(\mur),Q^2/\mur^2,Q^2/\muf^2) =  \nonumber \\ 
&\sigma_{a\bar{a}\to F}^{(0)}(\as(Q)) H^F(\as(Q))\left(\tilde{C}(N,\as(Q))\right)^2 \exp\left\{\int_{\muf^2}^{Q^2}\frac{dq^2}{q^2}2\gamma(N,\as(q))\right\},
\end{align}
and $\gamma(N,\as)$ is the DGLAP anomalous dimension. Finally the Sudakov exponent is given by
\begin{equation}\label{eq:sudakov-qt}
\Gq(N,\as(\mur),b,Q^2/\mur^2)=-\int_{Q^2/{\bar b}^2}^{Q^2}\frac{dq^2}{q^2}\left[A(\as(q))\log\frac{Q^2}{q^2}+\tilde{B}(N,\as(q))\right],
\end{equation}
with
\begin{equation}\label{eq:mod-B}
\tilde{B}(N,\as)=B(\as)+2\beta(\as)\frac{d\log \tilde{C}(N,\as)}{d\log\as}+2\gamma(N,\as),
\end{equation}
where the running coupling renormalization group equation is defined as
\begin{align}
 \beta(\as)=-\as \sum_{k=0}^\infty\left( \frac{\as}{\pi}\right)^{k+1} \beta_k.
\end{align}
We will be restoring the full flavor dependence for our numerical studies.
The treatment of the exponentiation of the full flavor dependence is explained in detail in Appendix A of~\cite{Bozzi:2005wk}. This method is implemented in the computer code \DYqt~\cite{Bozzi:2008bb,Bozzi:2010xn}, which we employ in our numerical studies.

\section{Combining transverse momentum and threshold resummations}\label{sec:combination}
We write the  transverse-momentum distribution in joint resummation as~\cite{Kulesza:2002rh}
\begin{equation}\label{hadron-level-joint}
\frac{d\sigma_{F}^{\mathrm{(res)}}}{dQ_T^2} = \int_{0}^{\infty}db\frac{b}{2}J_0(bQ_T)\int_{C_{T}}\frac{dN}{2\pi i} \left(\frac{Q^2}{s}\right)^{-N+1}\tilde{W}^F_\text{joint}(b,N,Q),
\end{equation}
where $\tilde{W}^F_\text{joint}(N,b,Q)$ has been defined in analogy with the function $\tilde{W}^F(N,b,Q)$ that appears in $Q_T$ resummation Eq.~(\ref{WqT})
\begin{align}\label{eq:WF-joint}
\tilde{W}^F_\text{joint}(N,b,Q) &=\; \mathcal{H}^{F}_\text{joint}(N,Q,\as(\mur),Q^2/\mur^2,Q^2/\muf^2) \;\tilde{f}_{a/h_1}(N,\muf^2)\;\tilde{f}_{\bar a/h_2}(N,\muf^2)\nonumber\\
&\times \;\exp\left\{\Gj(N,\as(\mur),b,Q^2/\mur^2)\right\}.
\end{align}
The aim of this section is to revise in some detail the calculation of the Sudakov exponent $\Gj$ and hard factor $\mathcal{H}^{F}_\text{joint}$ so that Eq.~(\ref{hadron-level-joint}) is valid at NLL accuracy. The extension to NNLL will be instead discussed in Section~\ref{sec:nnll}.

In order to provide a better understanding concerning the origin of the $Q_T$ and threshold logarithms, and their overlap, we study the $\Ord(\as)$ correction to the DY process in the eikonal approximation. We consider the emission of a soft (real) gluon off a quark-antiquark dipole. Because we are seeking an all-order result, we consider a two-dimensional Fourier transform with respect to the soft gluon transverse momentum, as well as a Laplace transform with respect to the gluon energy~\footnote{Laplace moments with respect to $2k_0\propto(1-z)$ are equivalent at leading power to the more familiar Mellin moments with respect to $z$, see e.g. Ref.~\cite{Laenen:2000ij}.}. We work in $d=4-2\epsilon$ dimensions and employing the $\overline{\text{MS}}$ scheme, we obtain:
\begin{align}
\Gj^\text{NLO}\left(b,N\right) &=   8\pi \as  C_{F}\left(\frac{\mu^2e^{\gammae}}{4\pi}\right)^\epsilon\int\frac{d^{4-2\epsilon}k}{\left(2\pi\right)^{3-2\epsilon}}\theta\left(k_{0}\right)\delta\left(k^{2}\right) e^{-\frac{2k_{0}}{Q}N-i\mathbf{b}\cdot\mathbf{k_T}}\,\frac{p_1\cdot p_2}{(p_1\cdot k)(p_2\cdot k)}
\nonumber\\
&= 8\pi  \as C_{F}\left(\frac{\mu^2e^{\gammae}}{4\pi}\right)^\epsilon\int\frac{d^{4-2\epsilon}k}{\left(2\pi\right)^{3-2\epsilon}}\theta\left(k_{0}\right)\delta\left(k^{2}\right) e^{-\frac{k_{+}+k_{-}}{Q}N-i\mathbf{b}\cdot\mathbf{k_T}}\, \frac{2}{k_{T}^{2}},
\end{align}
where in the second line we have made use of light-cone coordinates and set $|\mathbf{k_T}|^2=k_T^2$.
The integral over the light-cone components $k_-$ and $k_+$ can be easily performed, leading to
\begin{equation}\label{eq:NLO-eik}
\Gj^\text{NLO}\left(b,N\right) =  8\pi \as C_{F}\left(\frac{\mu^2e^{\gammae}}{4\pi}\right)^\epsilon\int\frac{d^{2-2\epsilon}k_T}{\left(2\pi\right)^{3-2\epsilon}} e^{-i\mathbf{b}\cdot\mathbf{k_T}}\frac{2}{k_{T}^{2}}\left[K_0\left(\frac{2Nk_T}{Q}\right)+\Ord \left(\frac{k_T^2}{Q^2} \right)\right],
\end{equation}
where $K_0$ is the modified Bessel function of the second kind of order zero. In order to remove the infrared divergencies from this real emission contribution, the virtual corrections and the contribution from the PDFs also need to be included. We obtain
\begin{equation}
\Gj^\text{NLO}\left(b,N\right) =   8\pi \as C_{F}\left(\frac{\mu^2e^{\gammae}}{4\pi}\right)^\epsilon\int\frac{d^{2-2\epsilon}k_T}{\left(2\pi\right)^{3-2\epsilon}} \frac{2}{k_{T}^{2}}\left[e^{-i\mathbf{b}\cdot\mathbf{k_T}}K_0\left(\frac{2Nk_T}{Q}\right)-\log\left(\frac{Q}{k_T}\right)+\log\bar N\right],
\end{equation}
where the second term is the result of the virtual contribution and can be identified as the integral of $1/(2k_+)$ over $k_+$. The final term is the result of the PDF contribution with $\bar N = N e^{\gammae}$. Combining the logarithms and integrating over the azimuthal angle results in:
\begin{equation}\label{eq:NLO-joint-epsi}
\Gj^\text{NLO}\left(b,N\right) =  2\frac{\as}{\pi}C_{F}\left(\mu^2e^{\gammae}\right)^\epsilon\int_0^{Q^2}\frac{dk_T^2}{k_T^{2+2\epsilon}} \left[\left(\frac{bk_T}{2}\right)^{\epsilon}J_{-\epsilon}\left(bk_T\right)K_0\left(\frac{2Nk_T}{Q}\right)+\frac{1}{\Gamma(1-\epsilon)}\log\left(\frac{\bar N k_T}{Q}\right)\right],
\end{equation}
where $J_a$ is the modified Bessel function of the first kind of order $a$.
The logarithm and the $K_0$ Bessel function cancel one another in the limit $k_T\to 0$, because $K_0\left(x\right)\sim -\log\left(xe^{\gammae}/2\right)$ and $J_0\sim1$. Due to this cancellation, the integral is infrared finite and we can evaluate it in $\epsilon\to0$ limit:
\begin{equation}\label{eq:NLO-joint}
\Gj^\text{NLO}\left(b,N\right) =  2\frac{\as}{\pi}C_{F}\int_0^{Q^2}\frac{dk_T^2}{k_T^{2}} \left[J_{0}\left(bk_T\right)K_0\left(\frac{2Nk_T}{Q}\right)+\log\left(\frac{\bar N k_T}{Q}\right)\right].
\end{equation}
This calculation can be extended to all orders and it leads to the simultaneous resummation of logarithms of $N$ and $b$, as discussed in detail in Ref.~\cite{Laenen:2000ij}. The resummed exponent at NLL can be written in a rather compact form:
\begin{align}
\Gj^\text{NLL}\left(b,N,Q,\mu_F\right) =& 2 \int_0^{Q^2}\frac{dq^2}{q^2}\left\{ A(\as(q)) \left[J_0\left(bq\right)K_0\left(\frac{2Nq}{Q}\right)+\log\left(\frac{\bar N q}{Q}\right)\right]\right\}\nonumber\\
& -2\log\bar N \int_{\muf^2}^{Q^2}\frac{dq^2}{q^2} A(\as(q)).
\end{align}
Note that the first term is essentially the running-coupling generalization of the one-loop computation in Eq.~(\ref{eq:NLO-joint}), while the second term takes into account the difference between the factorization scale $\mu_F$ and the hard scale of the process $Q$. The NLL result above can be further manipulated and rewritten a way that is similar to both $Q_T$ and threshold resummation and it is suitable for phenomenological applications~\cite{Kulesza:2002rh}.  
In particular, we use the fact that up to NNLL accuracy we can replace the Bessel function $J_0$ with a step function (see Appendix~\ref{sec:J0} for details): $J_{0}\left(bq\right)\to 1-\Theta\left(q-Q/\bar{b}\right)=\Theta\left(Q/\bar{b}-q\right)$. Thus we obtain
\begin{align}
\Gj^\text{NLL}\left(b,N,Q,\mu_F\right) = & 2\int_{0}^{Q^{2}/\bar{b}^{2}}\frac{dq^{2}}{q^{2}}A\left(\alpha_{\mathrm{s}}\left(q\right)\right)K_{0}\left(\frac{2Nq}{Q}\right)\nonumber\\&+
2\int_{0}^{Q^{2}}\frac{dq^{2}}{q^{2}} A\left(\alpha_{\mathrm{s}}\left(q\right)\right)\log\left(\frac{\bar{N}q}{Q}\right)
 -2\log\bar{N}\int_{\mu_{F}^{2}}^{Q^{2}}\frac{dq^{2}}{{q}^{2}}A\left(\alpha_{\mathrm{s}}\left(q\right)\right).
\end{align}
Following Refs.~\cite{Laenen:2000ij,Kulesza:2002rh} we note that the desired logarithmic behavior is capture if the Bessel function $K_{0}\left(x\right)$ is expanded at small values of its argument, provided that the upper bound of the integration is changed from $Q^{2}/\bar{b}^{2}$
to $Q^{2}/\chi^{2}$
\begin{align}\label{eq:Eeik-joint}
\Gj^\text{NLL}\left(\chi,N,Q,\mu_F\right) = &- 2\int_{0}^{Q^{2}/\chi^2}\frac{dq^{2}}{q^{2}}A\left(\alpha_{\mathrm{s}}\left(q\right)\right)\log\left(\frac{Nq}{Q}\right)\nonumber\\&+
2\int_{0}^{Q^{2}}\frac{dq^{2}}{q^{2}} A\left(\alpha_{\mathrm{s}}\left(q\right)\right)\log\left(\frac{\bar{N}q}{Q}\right)
 -2\log\bar{N}\int_{\mu_{F}^{2}}^{Q^{2}}\frac{dq^{2}}{q^{2}}A\left(\alpha_{\mathrm{s}}\left(q\right)\right) \nonumber\\
 &=  2\int_{Q^2/\chi^2}^{Q^2}\frac{dq^2}{q^2}A(\as(q)) \log\left(\frac{\bar{N}q}{Q}\right) -2\log\bar N \int_{\muf^2}^{Q^2}\frac{dq^2}{q^2}A(\as(q)).
\end{align}
The function $\chi(\bar{N},\bar{b})$ is defined so that it behaves as $\bar{b}$ in the large $\bar{b}$ limit and as $\bar{N}$ in the large $\bar{N}$ limit. Furthermore, if we require $\chi(\bar{N},0)=\bar{N}$, then the integral over $Q_T$ results in the inclusive threshold-resummed cross section. An example of such a function is $\chi=\bar{b}+\bar{N}$. 
For $\bar b=0$, $\chi= \bar N$ and Eq.~(\ref{eq:Eeik-joint}) reduces to the threshold Sudakov exponential up to NLL accuracy. 
We can also re-arrange the contributions in a different way, so that the result resembles more closely the Sudakov exponent that appears in $Q_T$ resummation:
\begin{align}\label{eq:Sudakov-joint}
\Gj^\text{NLL}\left(\chi,N,Q,\mu_F\right) =&  -\int_{Q^2/\chi^2}^{Q^2}\frac{dq^2}{q^2}\left[A(\as(q)) \log\left(\frac{Q^2}{q^2}\right)+B(\as(q))\right]\nonumber\\
& +\int_{\muf^2}^{Q^2/\chi^2}\frac{dq^2}{q^2}\left[-2\log\bar N A(\as(q))-B(\as(q))\right].
\end{align}
The first term can be recognized as the exponential for $Q_T$ resummation Eq.~(\ref{eq:sudakov-qt}), with the replacement $\bar b \to \chi$, and the second term is the large $\bar N$ limit of the DGLAP evolution of the PDFs from a scale $Q/\chi$ to $\mu_F$. 
However, because the identification of $B^{(i)}$ with the constant part of the DGLAP anomalous dimension (i.e. the $\delta$-function contribution to the splitting function) only holds for $B^{(1)}$, this way of rewriting the joint Sudakov exponent only holds up to NLL accuracy. The extension to NNLL accuracy will be discussed in the next section. 
It is worth pointing out a difference in the logarithmic counting between joint and transverse momentum resummation. DGLAP contributions affect the $Q_T$ spectrum with single logarithms of $Q_T$. However, the flavor-diagonal anomalous dimensions carry an additional $A(\as) \log \bar{N}$ contribution. Therefore, when computing joint resummation at N$^k$LL order, parton evolution, or at least its large-$N$ behavior, has to be included up to N$^k$LO order, while N$^{k-1}$LO suffices for $Q_T$.

At NLL level the treatment of the hard factor is relatively straightforward because the one-loop coefficient functions $\tilde{C}$ do not contain logarithms of $N$, i.e. $D^{(1)}=0$. However, we have to make sure that the threshold-enhanced part of the $\muf$-dependent contribution is exponentiated. We have
\begin{align}
\Gj^\text{NLL}(\as(\mur),b,N,Q^2/\mur^2)=&-\int_{Q^2/\chi^2}^{Q^2}\frac{dq^2}{q^2}\left[A(\as(q))\log\frac{Q^2}{q^2}+\tilde{B}(N,\as(q))\right]
+2\int_{\muf^2}^{Q^2}\frac{dq^2}{q^2}\gamma(N,\as(q)),
\end{align}
which is equivalent to Eq.~(\ref{eq:Sudakov-joint}) in the large $N$ limit, and the hard factor is simply
\begin{equation}
\mathcal{H}^{F,\text{NLL}}_\text{joint}(N,Q,\as(\mur),Q^2/\mur^2,Q^2/\muf^2) =\;\sigma_{a\bar{a}\to F}^{(0)}
\left[1+ \frac{\as(Q)}{\pi}\left(H^{F,(1)}+ 2 C^{(1)}(N)\right) \right].
\end{equation}
We will see in the next section that in order to achieve NNLL accuracy, the way we treat $\mathcal{H}^{F}$ must be refined.

\section{Joint resummation at NNLL} \label{sec:nnll}
After recalling the main ingredients of joint resummation, we are ready to implement it to NNLL accuracy. We discuss first the resummed exponent, followed by an analysis of the hard factor.

\subsection{Sudakov exponent at NNLL}
A few issues must be addressed in order to ensure NNLL accuracy in both $N$ and $b$. Firstly, the full PDF evolution now needs to be taken into account at NLO accuracy, together with the large $N$ limit of the NNLO anomalous dimension. At the central scale the latter is computed as:
\begin{equation}
\Delta \Gj^{\mathrm{DGLAP}}=-2 A^{(3)}_\text{thr}\log\bar N \int_{\muf^2}^{Q^2/\chi^2}\frac{dq^2}{q^2} \left(\frac{\as(q)}{\pi}\right)^3.
\end{equation}
A second term that starts to contribute at NNLL accuracy is the soft wide-angle contribution of threshold resummation, $\as^2\tilde{D}^{(2)}\log \bar N$. This contribution is not exponentiated in $Q_T$ resummation but, it is present in $C^{(2)}_{aa}$ Eq.~(\ref{QTcoeff}), or equivalently in $\mathcal{H}^{(2)}$ Eq.~(\ref{HfuncQT}), while it contributes to the Sudakov exponent of threshold resummation Eq.~(\ref{eq:sudakov-thr})
\begin{equation}
\Delta \Gj^\mathrm{wide-angle}=-\frac{1}{2}\int^{Q^2}_{Q^2/\bar{N}^2}\frac{dq^2}{q^2}\tilde{D}(\as(q))
\end{equation}
Thus, this term contributes to the resummed exponent in joint resummation and it has to be subtracted from $\mathcal{H}^{(2)}$ in order to prevent double counting:
\begin{equation}
\mathcal{H}^{(2)}\to\mathcal{H}^{(2)}+\tilde{D}^{(2)}\log\bar{N}.
\end{equation}
A similar method was performed for joint resummation of heavy quark production, where the soft wide-angle contribution enters at NLL \cite{Banfi:2004xa}.
The last contribution to the exponent that we need to take into account is the aforementioned difference between $A_{Q_T}^{(3)}$ and $A_{\mathrm{thr}}^{(3)}$. 
\begin{equation}\label{eq:col-anom}
\Delta \Gj^{\mathrm{cusp}}=-\int_{Q^2/\chi^2}^{Q^2/\bar{N}^2}\frac{dq^2}{q^2}\left(A^{(3)}_{Q_{T}}-A^{(3)}_{\mathrm{thr}}\right)\left(\frac{\as(q)}{\pi}\right)^{3}\log\frac{Q^2}{q^2},
\end{equation}
where $A^{(3)}_{Q_{T}}-A^{(3)}_{\mathrm{thr}}=-\beta_0 \tilde{D}^{(2)}$. 
In the language of SCET this contribution is known as the collinear anomaly~\cite{Becher:2010tm}. 
It  essentially arises because one evaluates both soft and collinear contributions at the same scale~\cite{Monni:2011gb}. 
Note that we have some freedom in choice of the integration boundaries in Eq.~(\ref{eq:col-anom}). We demand that the lower limit approaches $Q^2/ \bar b^2$ in the large $\bar b$ limit, in order to reproduce the $Q_T$ case. Moreover, with the above choice this contribution vanishes the inclusive case $\chi(\bar N, 0)=\bar N$. 

\subsection{Treatment of the hard factor}

Thus far we have concentrated on discussing the Sudakov exponent. However, the pre-factors for $Q_T$ and threshold resummation, $\mathcal{C}$ and $\mathcal{H}$, respectively in Eq.~(\ref{CfuncTH}) and Eq.~(\ref{HfuncQT}), actually differ already at one-loop level (see Ref.~\cite{Catani:2013tia} for an all-order discussion). In order to better understand this difference, it is useful to go back to the one-loop calculation of Section~\ref{sec:combination}. 
In particular, we can perform the transverse momentum integral at fixed-coupling in Eq.~(\ref{eq:NLO-joint-epsi}), keeping the upper limit of the integration for the virtual and for the collinear counterterm at $Q$ and taking $\epsilon \to 0$, we find (see also~\cite{Li:2016axz})
\begin{equation}
\Gj^\text{NLO}\left(b,N\right)=\frac{\as}{\pi} C_{F}\left[2\log^{2}\bar{N}+{\rm Li}_{2}\left(-\frac{\bar{b}^{2}}{\bar{N}^{2}}\right)+\zeta_{2}\right].
\end{equation}
If this is approximated in the threshold limit, i.e. at large $N$, it approaches
\begin{equation}
\lim_{N\to\infty}\Gj^\text{NLO}\left(b,N\right)=\frac{\as}{\pi}C_{F}\left[\zeta_{2}+2\log^{2}\bar{N}\right].
\end{equation} 
If instead this is approximated in the $Q_T$-resummation limit, large $b$, this results in
\begin{equation}
\lim_{b\to\infty}\Gj^\text{NLO}\left(b,N\right)=\frac{\as}{\pi}C_{F}\left[-2\log^{2}\bar{b}+4\log\bar{b}\log\bar{N}\right].
\end{equation}
This reproduces the logarithmic structure of both threshold and $Q_T$ resummation, however the constant term has a difference of $\zeta_2=\pi^2/6$, which is indeed the difference between the $\mathcal{H}$ and $\mathcal{C}$ at one-loop~\cite{Kulesza:2002rh,Catani:2013tia}. In order to account for this difference we add the NLO computation minus the expansion of the logarithmic exponential at NLO:
\begin{align}\label{eq:DeltaH}
\Delta{\cal H}^{\left(1\right)} = &\; A^{(1)}\left[2\log^{2}\bar{N}+{\rm Li}_{2}\left(-\frac{\bar{b}^{2}}{\bar{N}^{2}}\right)+\zeta_{2}+2\log^{2}\chi-4\log\chi\log\bar{N}\right]\nonumber \\
= &\; A^{(1)}\left[\zeta_{2}+{\rm Li}_{2}\left(-\frac{\bar{b}^{2}}{\bar{N}^{2}}\right)+2\log^{2}\left(\chi/\bar{N}\right)\right]
\simeq A^{(1)}\left[\zeta_{2}-{\rm Li}_{2}\left(\frac{\bar{b}^{2}}{\chi^2}\right)\right],
\end{align}
where the last step is valid up to power-suppressed terms.
There is also an analogous term in $\mathcal{H}^{(2)}$, the correct treatment of which would be necessary in order to achieve NNLL$^\prime$ accuracy in both threshold and $Q_T$. However, in this work we only consider NNLL accuracy for joint resummation and therefore we do not have to worry about it. For our numerical studies we take this contribution from transverse resummation and, therefore, we do reach NNLL$^\prime$ for $Q_T$ but not for threshold resummation.

Note that the modification of $\mathcal{H}^{(1)}$ not only influences the hard coefficient, but also the exponential. The $N$-dependent contribution that we find is naturally a part of $C_{aa}$ and therefore it should be computed with the strong coupling $\as$ at the scale $Q/\chi$. If we then express the pre-factor at the hard scale, we induce a new term in the resummed exponent, see Eq.~(\ref{eq:mod-B}), which effectively amounts to a modification of the coefficient $\tilde{B}^{(2)}$ :
\begin{equation} \label{newB2}
\tilde{B}^{(2)} \to\tilde{B}^{(2)}-\beta_0 \Delta{\cal H}^{\left(1\right)}.
\end{equation}

\subsection{NNLL joint cross section}\label{NNLL-joint}
We are now ready to put together all the contributions discussed in the previous sections and finally arrive at an expression for the DY transverse momentum spectrum that simultaneously resums threshold and $Q_T$ logarithms to NNLL.
We start with the resummed exponent, that reads
\begin{align}\label{eq:joint-exponent}
\Gj^{\mathrm{NNLL}}(\as(\mur),b,N,Q^2/\mur^2)=&-\int_{Q^2/\chi^2}^{Q^2}\frac{dq^2}{q^2}\left[A(\as(q))\log\frac{Q^2}{q^2}+\tilde{B}(N,\as(q))\right]\nonumber\\
&\hspace{-5em}+\int_{\muf^2}^{Q^2}\frac{dq^2}{q^2}2\gamma(N,\as(q))-2A_\text{thr}^{(3)}\log\bar N\int_{\muf^2}^{Q^2/\chi^2}\frac{dq^2}{q^2}  \left(\frac{\as(q)}{\pi}\right)^3\nonumber\\
&\hspace{-5em}-\frac{1}{2}\int^{Q^2}_{Q^2/\bar{N}^2}\frac{dq^2}{q^2}\tilde{D}_{a}(\as(q))+\beta_0A^{(1)}\left[\zeta_{2}-{\rm Li}_{2}\left(\frac{\bar{b}^{2}}{\chi^2}\right)\right]\int_{Q^2/\chi^2}^{Q^2}\frac{dq^2}{q^2}\left(\frac{\as(q)}{\pi}\right)^2\nonumber\\
&\hspace{-5em}-\int_{Q^2/\chi^2}^{Q^2/\bar{N}^2}\frac{dq^2}{q^2}\left(A^{(3)}_{Q_{T}}-A^{(3)}_{\mathrm{thr}}\right)\left(\frac{\as(q)}{\pi}\right)^{3}\log\frac{Q^2}{q^2}.
\end{align}
Note the second line of this expression contains the contribution from DGLAP evolution: the anomalous dimension $\gamma$ is taken up to NLO order, while its NNLO contribution is considered only in the soft limit. The above result can be brought to a rather compact form (details are given in Appendix~\ref{sec:coef}):
\begin{align}\label{eq:joint-exponent-final}
\Gj^{\mathrm{NNLL}}(\as(\mur),b,N,Q^2/\mur^2)=& -\int_{Q^2/\chi^2}^{Q^2}\frac{dq^2}{q^2}\Bigg[A(\as(q))\log\frac{Q^2}{q^2}+\tilde{B}(N,b,\as(q))+\frac{1}{2}\tilde{D}(\as(q))\Bigg]\nonumber \\&- \frac{1}{2}\log \left(\frac{\bar N^2}{\chi^2}\right) \tilde{D}\left(\as\left(\frac{Q}{\chi}\right) \right)+2\int_{\muf^2}^{Q^2}\frac{dq^2}{q^2}\gamma_\text{soft}(N,\as(q)),
\end{align}
where now $A$ is always the cusp and $\tilde{B}(N,b,\as)$ has been put into a form that closely resembles the analogous coefficient $\tilde{B}(N,\as)$ appearing in $Q_T$ resummation Eq.~(\ref{eq:mod-B}):
\begin{equation}
\tilde{B}(N,b,\as)=B(\as) +2 \beta(\as)\frac{d\log \tilde{C}(N,b,\as)}{d\log \as}+2\gamma(N,\as).
\end{equation}
Note however that the coefficient function $\tilde{C}$ differs in two ways with respect to the one entering standard $Q_T$ resummation: first, threshold-enhanced terms are subtracted off and, secondly, it contains the contribution
\begin{align}
F(N,b,\as)= \frac{\as}{\pi} A^{(1)} \left[\zeta_{2}-{\rm Li}_{2}\left(\frac{\bar{b}^{2}}{\chi^2}\right)\right] + \mathcal{O}\left(\as^2\right),
\end{align}
resulting in
\begin{equation}
\tilde{C}(N,b,\as)=\tilde{C}(N,\as)+\frac{F(N,b,\as)}{2}+\frac{1}{2}\left(\frac{\as}{\pi}\right)^2\tilde{D}^{(2)}\log\bar{N},
\end{equation}
where $\tilde{C}(N,\as)$ is the same as for $Q_T$ resummation (see Appendix~\ref{sec:coef} for explicit expressions for the coefficients). In order to achieve NNLL in both variables, the DGLAP anomalous dimension $\gamma$ needs to be evaluated at $n$NLO accuracy, which is defined as NLO accuracy plus the $\log \bar N$ contributions from the NNLO. 
On the other hand, $\gamma_\text{soft}$ in Eq.~(\ref{eq:joint-exponent-final}) only contains the threshold-enhanced, while the residual $\muf$ dependence is included at fixed-order. We note that Eq.~(\ref{eq:joint-exponent-final}) can easily be reduced to the threshold exponent by setting $\bar b=0$, i.e. $\chi=\bar N$. In order to recover the $Q_T$-resummation exponent in the limit $\chi \to \bar b$ a few algebraic steps are necessary, as detailed in Appendix~\ref{sec:coef}.
Finally, the hard factor is given by
\begin{equation}\label{eq:Hard-function}
\mathcal{H}^{F,\;\mathrm{NNLL}}_\text{joint}=1+\frac{\as}{\pi}\left\{\mathcal{H}^{F,(1)}+A^{(1)}\left[\zeta_{2}-{\rm Li}_{2}\left(\frac{\bar{b}^{2}}{\chi^2}\right)\right]\right\}+\left(\frac{\as}{\pi}\right)^2\left\{\mathcal{H}^{F,(2)}+\tilde{D}^{(2)}\log\bar{N}\right\}.
\end{equation}

Thus far, only the flavor diagonal contributions have been discussed in the context of joint resummation. However, the treatment of the full flavor dependence can be recovered by using the same method as for $Q_T$ resummation. The details of this method are described in Appendix A of~\cite{Bozzi:2005wk}. These off-diagonal contributions are suppressed in the threshold limit, therefore their inclusion in joint resummation comes with some freedom: we can can either include them or treat them only in $Q_T$ resummation, thus providing two results in joint resummation that differ by power-suppressed contributions in the threshold limit.

\section{Phenomenological studies}\label{sec:numerics}
Having obtained a joint resummed cross section at NNLL accuracy in both $Q_T$ and threshold, we can explore numerical results. 
In order to analyze the numerical effect of the joint resummation formalism we make use of a modified version of the \DYqt~code~\cite{Bozzi:2010xn,Bozzi:2008bb}. We also use  \DYqt~to produce results for $Q_T$ resummation only. We choose the CT14~\cite{Dulat:2015mca} set of parton distributions, which are used at NLO accuracy for the LO and LO+NLL$^\prime$ distributions and NNLO accuracy for NLO and NLO+NNLL. As is the case in~\cite{Kulesza:2002rh} we have chosen $\chi=\bar{b}+\bar{N}/(1+\eta\;\bar{b}/\bar{N})$ with the choice $\eta=1/4$. A more detailed discussion of the impact from different choices of $\chi$ or $\eta$ can be found in Appendix~\ref{sec:chi}. 
We first explore the expansion of the resummation and compare it to the fixed order computation. This allows us to comment on the validity of the approximation. Next, we present the fully resummed result at LO+NLL$^\prime$ and NLO+NNLL accuracies.

\subsection{Expansion}
\begin{figure}
	\centering
	\begin{tabular}{cc}
		\includegraphics[width=0.45\textwidth]{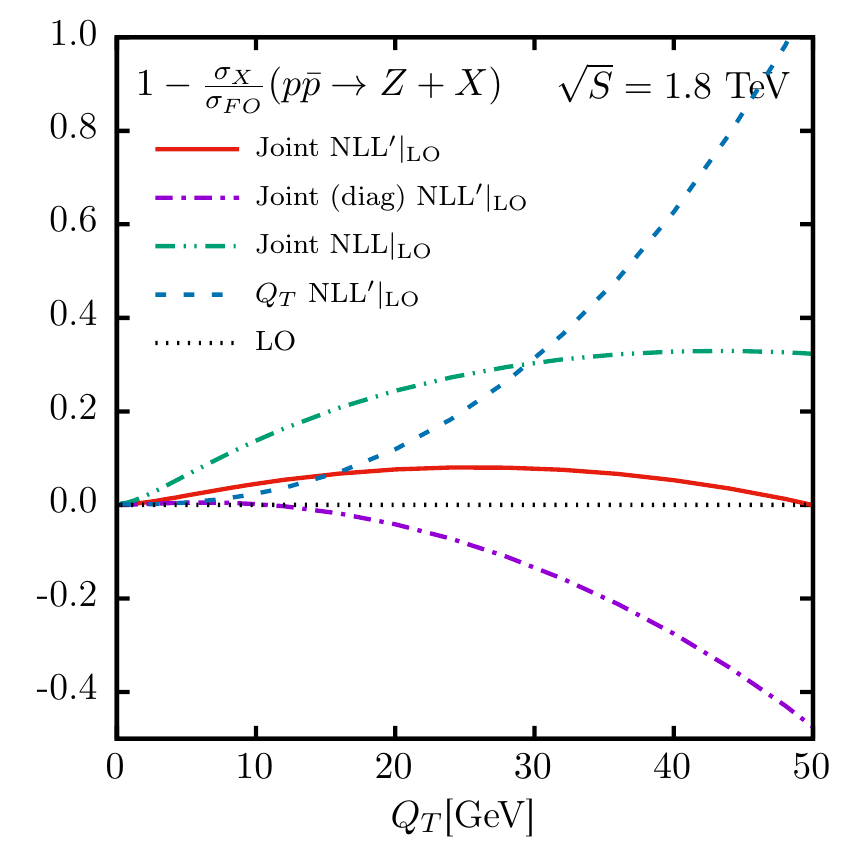} &
		\includegraphics[width=0.45\textwidth]{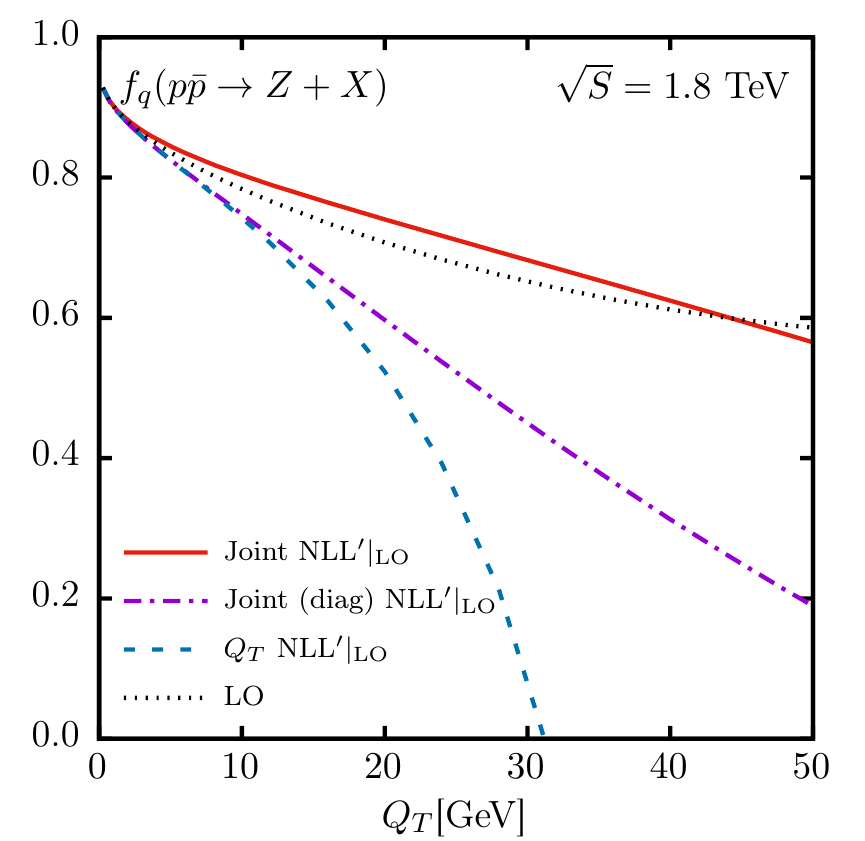} \tabularnewline
		\hspace{1.7em} (a) &\hspace{1.7em} (b) \tabularnewline
	\end{tabular}
	\caption{$Z$ boson transverse momentum distribution at the Tevatron with $\sqrt{S}=1.8$ TeV collision energy. The different approximations are obtained by expanding the NLL resummation to first order and they are compared to the LO result.
In the panel (a) the ratio to fixed order, $1-\frac{d\sigma_X}{dQ_T}/\frac{d\sigma_{\mathrm{LO}}}{dQ_T}$ is plotted, while in panel (b) the fraction in the $q\bar{q}$ channel, $f_q=\frac{d\sigma_q}{dQ_T}/\frac{d\sigma}{dQ_T}$ is shown.} 
	\label{fig:expansion-Teva-NLL}
\end{figure}

We start our study by considering the production of a $Z$ boson at the Tevatron, which is the same setup used in the previous study~\cite{Kulesza:2002rh}. In Fig.~\ref{fig:expansion-Teva-NLL} we show the comparison of the LO $Q_T$ distribution with the expansion of the joint and $Q_T$ resummation differential cross sections using different approximations.
The curve labelled "Joint~NLL$|_{\mathrm{LO}}$" corresponds to the expansion of the NLL result of Ref.~\cite{Kulesza:2002rh}, which does not include the modification of the hard coefficient. If the additional contribution to the hard coefficient is included, the lines indicated by~"Joint~NLL$^\prime|_{\mathrm{LO}}$" are obtained. 
Our default result corresponds to perform joint resummation also in the flavor off-diagonal contributions, which are usually not included in threshold resummation because they are power-suppressed (for recent progress on all-order understanding of power-suppressed contributions see Ref.~\cite{Bonocore:2016awd} and references therein).  This correctly captures the next-to-leading power corrections at $\Ord(\as)$, but provides only partial information beyond that.
Alternatively, one can exclude these contributions from joint resummation, so that the integral over $Q_T$ precisely reproduce the inclusive cross section obtained with threshold resummation, without additional power corrections. We implement this second resummation scheme by separating the contributions to the $\tilde{B}$ term in Eq.~(\ref{eq:joint-exponent-final}) in two classes: those that do not vanish at large $N$ are treated in joint resummation, while power-corrections in the threshold limit are only integrated over the range $[Q^2/\bar{b}^2,Q^2]$, which is the same as $Q_T$ resummation. 
The results of this second implementation are labeled "Joint~(diag)~NLL$^\prime|_{\mathrm{LO}}$", because they include only contribution from the $q\bar q$ initial state. We stress again that these two implementations of joint resummation are the same up to power corrections in the threshold limit.

\begin{figure}
	\centering
	\begin{tabular}{cc}
		\includegraphics[width=0.45\textwidth]{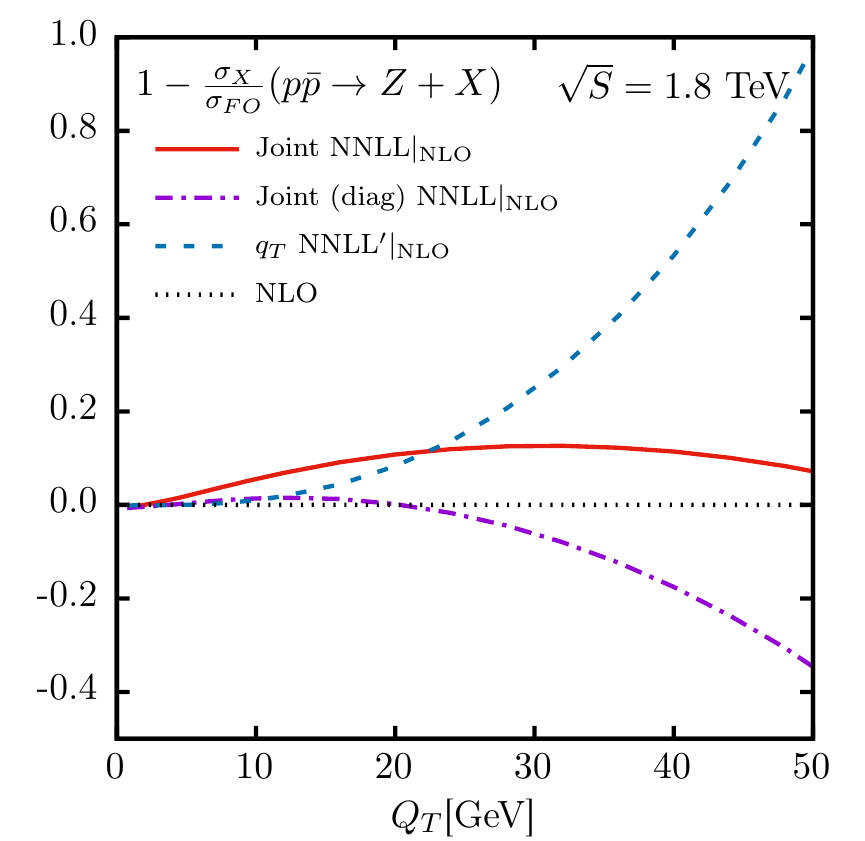} &
		\includegraphics[width=0.45\textwidth]{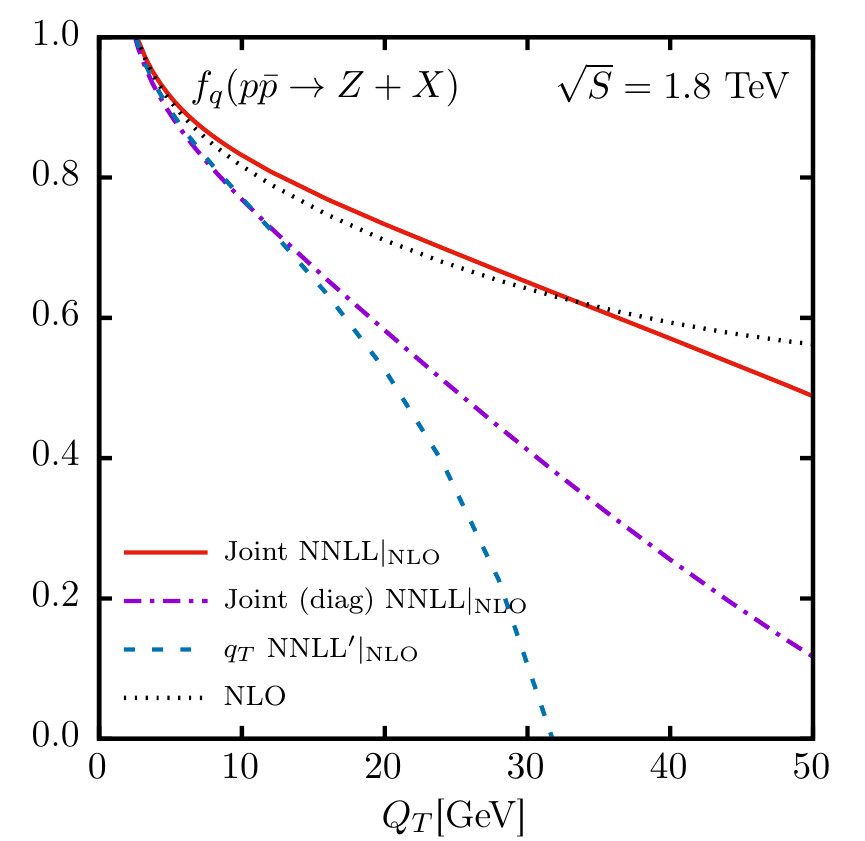} \tabularnewline
		\hspace{1.7em} (a) &\hspace{1.7em} (b) \tabularnewline
	\end{tabular}
	\caption{The same as Fig.~\ref{fig:expansion-Teva-NLL}, but comparing the expansion of the NNLL resummation to NLO accuracy.} 
	\label{fig:expansion-Teva-NNLL}
\end{figure}

In Fig.~\ref{fig:expansion-Teva-NLL}(a) we show $1-\frac{d\sigma_X}{dQ_T}/\frac{d\sigma_{\mathrm{LO}}}{dQ_T}$, where $X$ stands for the different approximations reported in the legend. From this plot we can see that the inclusion of additional threshold contribution to the hard coefficient produces a better agreement to the fixed order computation up to scales of at least 50 GeV.  Moreover, while the "Joint~(diag)" does not perform as well as our default implementation, it does better than just the expansion of $Q_T$ resummation.
In Fig.~\ref{fig:expansion-Teva-NLL}(b) we concentrate on the partonic subprocess that we have under theoretical control, namely $q \bar q$.
We plot the fraction of the cross section that can be attributed to the $q\bar{q}$ initial state channel: $f_q=\frac{d\sigma_q}{dQ_T}/\frac{d\sigma}{dQ_T}$. As we move to larger values of $Q_T$ the contribution from the other partonic channels become more significant and since these terms are not correctly approximated in the "Joint~(diag)" method this results in a deviation from the total fixed order differential cross section. Here it can also be seen that the $Q_T$ expansion is worse in this  individual channel, however a cancellation makes it work somewhat better for the sum of all channels. 
On the other hand, our default implementation for joint resummation does include power-suppressed contributions both in the $q \bar q $ and in the off-diagonal channels, which renders this type of cancellation more moderate.
In Fig.~\ref{fig:expansion-Teva-NNLL} a similar comparison can be seen at NLO accuracy in the $Q_T$ distribution. The conclusions are the same as for LO accuracy and joint resummation works just as well with the extension to NNLL accuracy as at NLL$^\prime$ accuracy.

\begin{figure}
	\centering
	\begin{tabular}{cc}
		\includegraphics[width=0.45\textwidth]{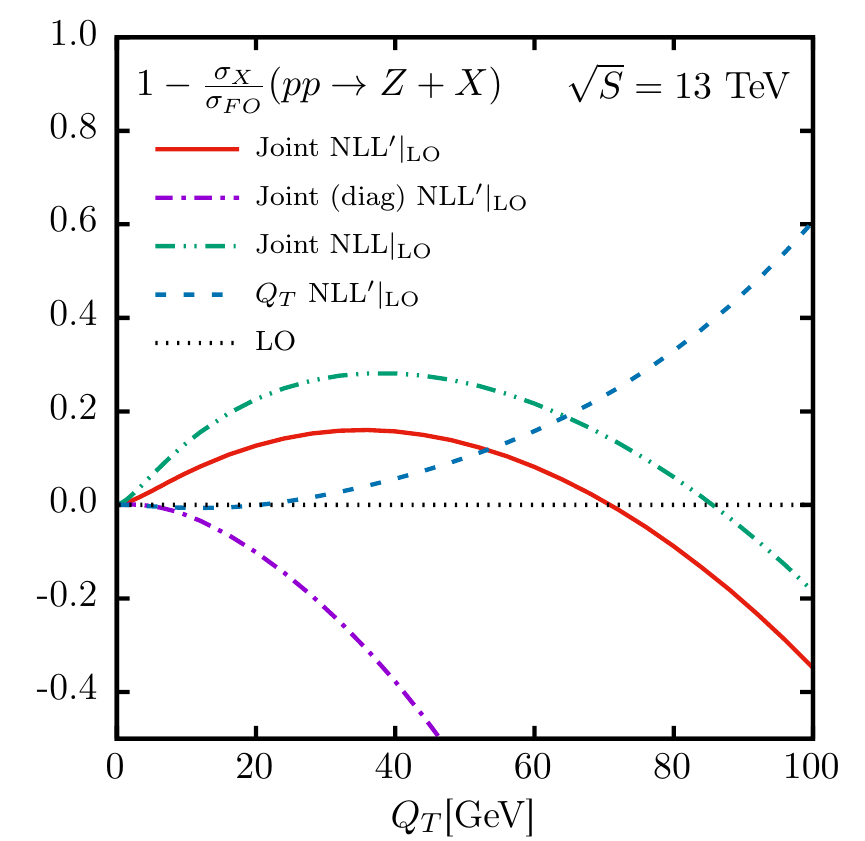} &
		\includegraphics[width=0.45\textwidth]{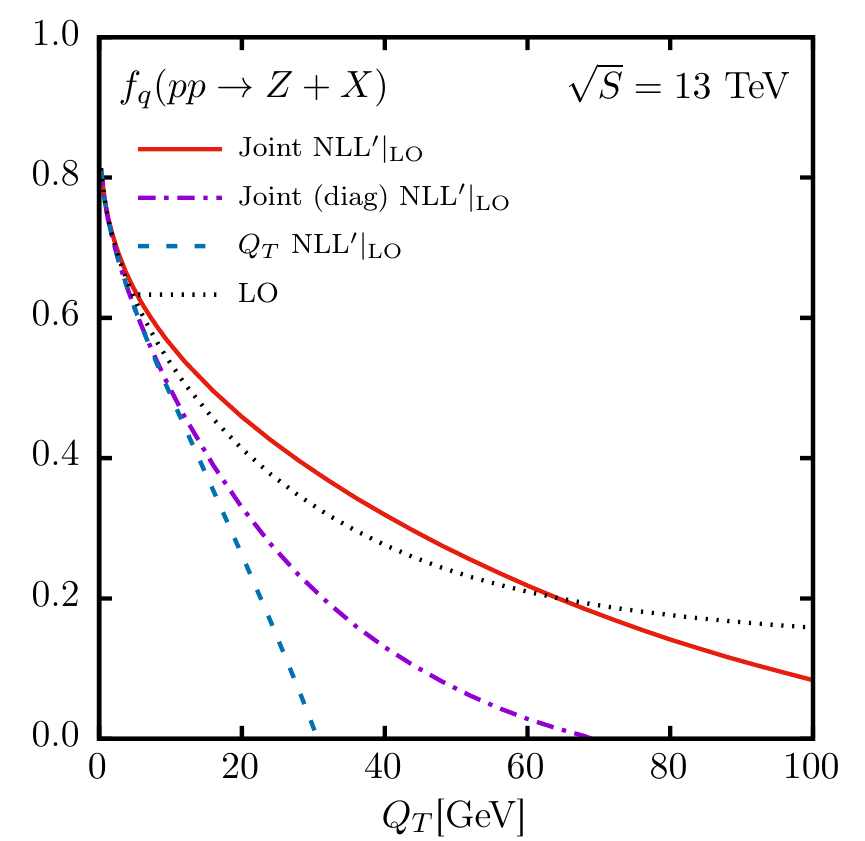} \tabularnewline
		\hspace{1.7em} (a) &\hspace{1.7em} (b) \tabularnewline
	\end{tabular}
	\caption{$Z$ boson transverse momentum distribution at the LHC with $\sqrt{S}=13$ TeV collision energy. The different approximations are obtained by expanding the NLL resummation to first order and they are compared to the LO result.
In the panel (a) the ratio to fixed order, $1-\frac{d\sigma_X}{dQ_T}/\frac{d\sigma_{\mathrm{LO}}}{dQ_T}$ is plotted, while in panel (b) the fraction in the $q\bar{q}$ channel, $f_q=\frac{d\sigma_q}{dQ_T}/\frac{d\sigma}{dQ_T}$ is shown.} 
	\label{fig:expansion-LHC-NLL}
\end{figure}
\begin{figure}
	\centering
	\begin{tabular}{cc}
		\includegraphics[width=0.45\textwidth]{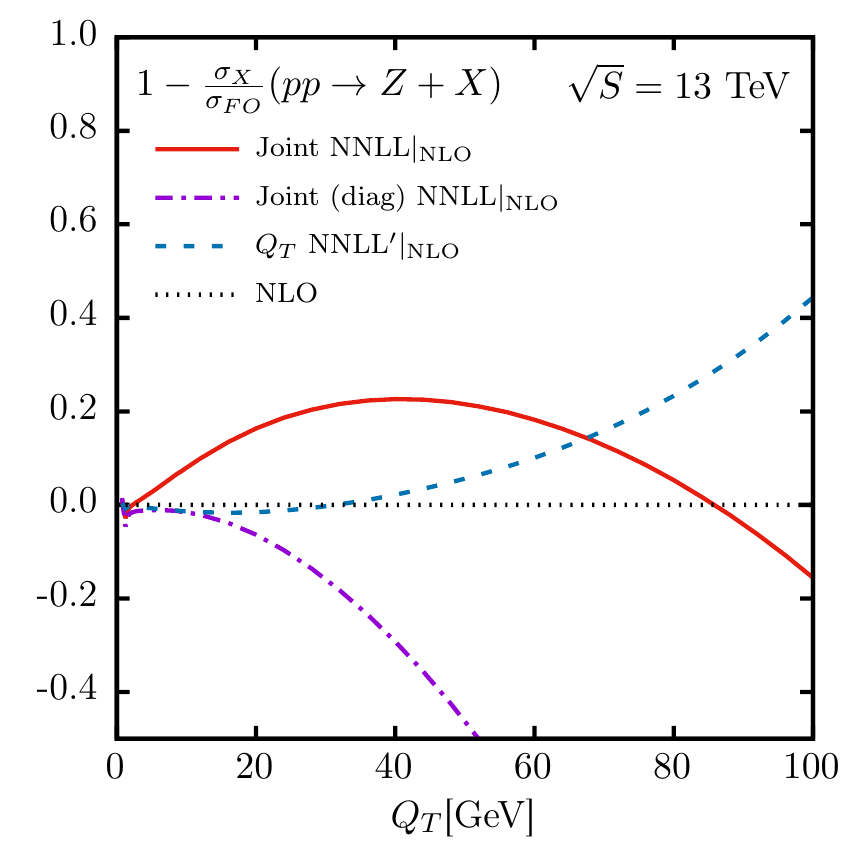} &
		\includegraphics[width=0.45\textwidth]{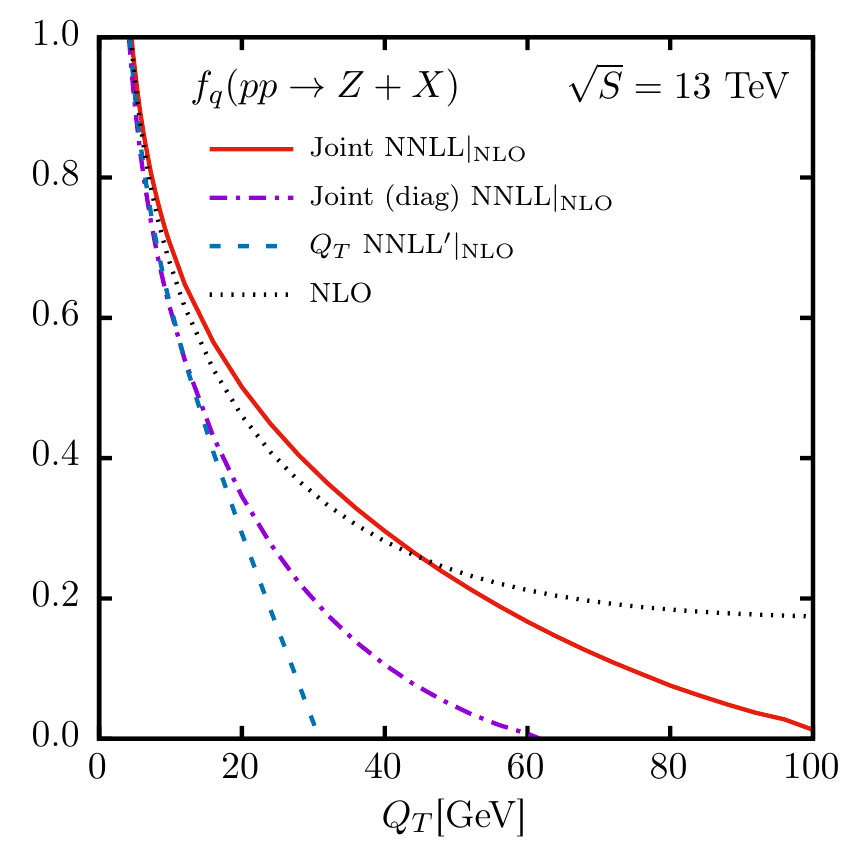} \tabularnewline
		\hspace{1.7em} (a) &\hspace{1.7em} (b) \tabularnewline
	\end{tabular}
	\caption{The same as Fig.~\ref{fig:expansion-LHC-NLL}, but comparing the expansion of the NNLL resummation to NLO accuracy.} 
	\label{fig:expansion-LHC-NNLL}
\end{figure}

We continue our study by considering $Z$ production the LHC at 13 TeV center-of-mass energy. 
Unfortunately, in this setup our findings are on less solid ground than what was obtained at Tevatron energies.
The results plotted in Fig.~\ref{fig:expansion-LHC-NLL}(a) prevent us to claim that the expansion of joint resummation provides an improved approximation of the fixed order over $Q_T$ resummation alone. This is perhaps surprising because if we look only at the $q \bar q$ channel, as in Fig.~\ref{fig:expansion-LHC-NLL}(b) we are drawn to the opposite, rather positive, conclusion. However, the same plot shows us that the importance of $q \bar q$ channel relatively to other partonic subprocesses is decreased. Moreover, power-corrections to the threshold expansion are rather important as indicated by the spread of our default and flavor-diagonal results.  This should not come as a surprise as we move further away from the threshold for $Z$ production.
Fig.~\ref{fig:expansion-LHC-NNLL} shows that the conclusions remain similar at the next perturbative order.

\begin{figure}
	\centering
	\begin{tabular}{cc}
		\includegraphics[width=0.45\textwidth]{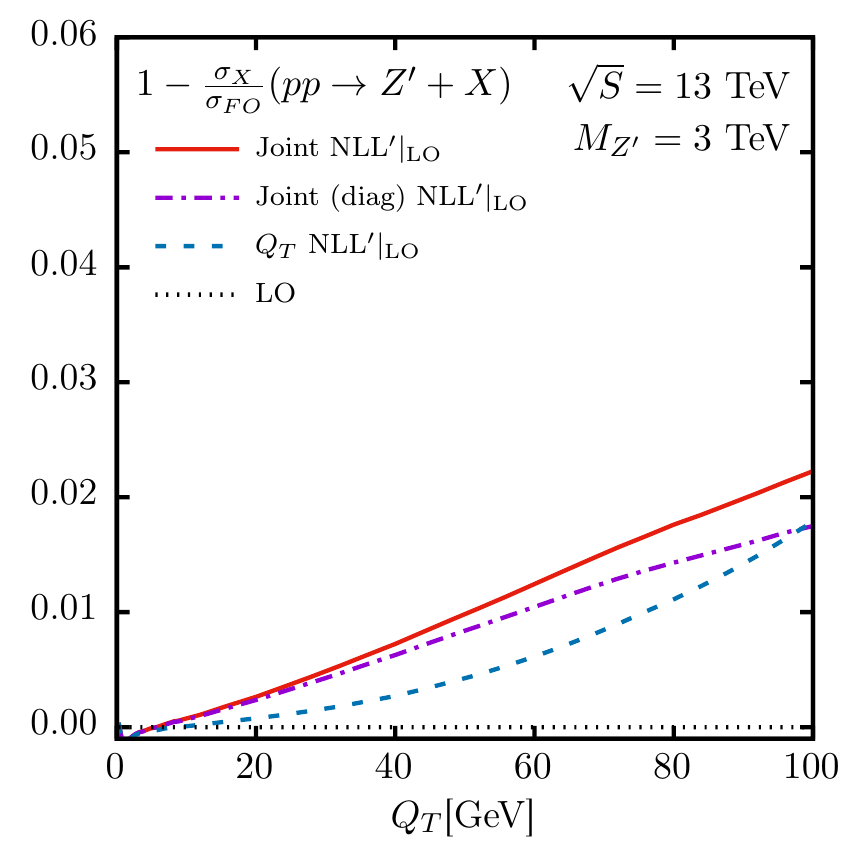} &
		\includegraphics[width=0.45\textwidth]{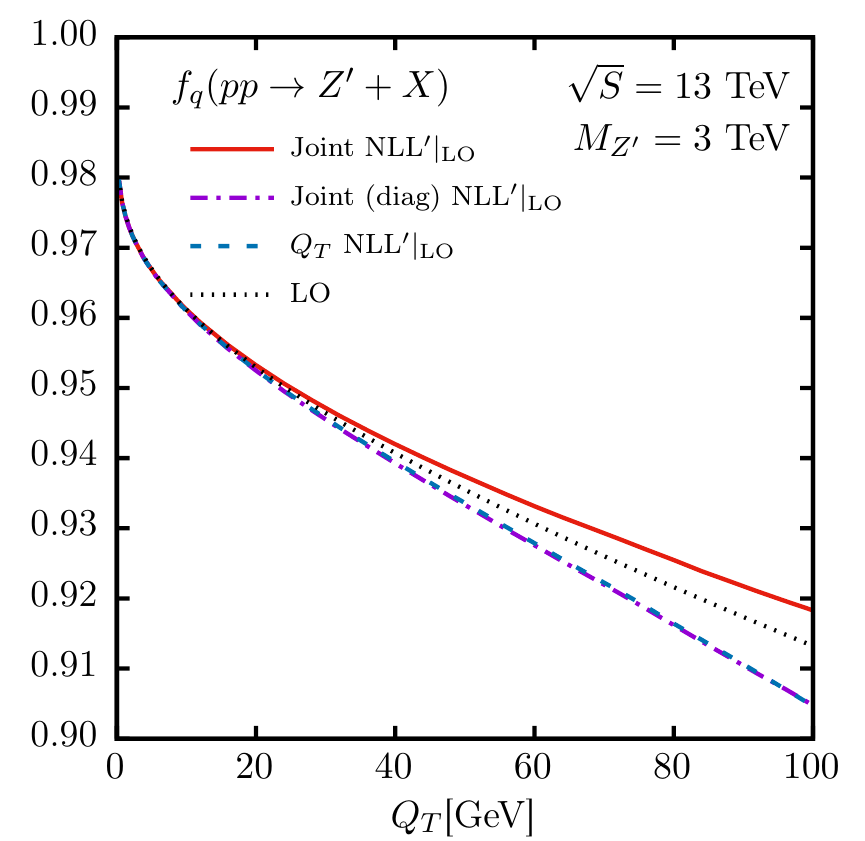} \tabularnewline
		\hspace{1.7em} (a) &\hspace{1.7em} (b) \tabularnewline
	\end{tabular}
	\caption{$Z^\prime$ boson transverse momentum distribution ($M_{Z^\prime}=3$ TeV) at the LHC with $\sqrt{S}=13$ TeV collision energy. The different approximations are obtained by expanding the NLL resummation to first order and they are compared to the LO result.
In the panel (a) the ratio to fixed order, $1-\frac{d\sigma_X}{dQ_T}/\frac{d\sigma_{\mathrm{LO}}}{dQ_T}$ is plotted, while in panel (b) the fraction in the $q\bar{q}$ channel, $f_q=\frac{d\sigma_q}{dQ_T}/\frac{d\sigma}{dQ_T}$ is shown.} 
	\label{fig:expansion-LHC-Zp-NLL}
\end{figure}

\begin{figure}
	\centering
	\begin{tabular}{cc}
		\includegraphics[width=0.45\textwidth]{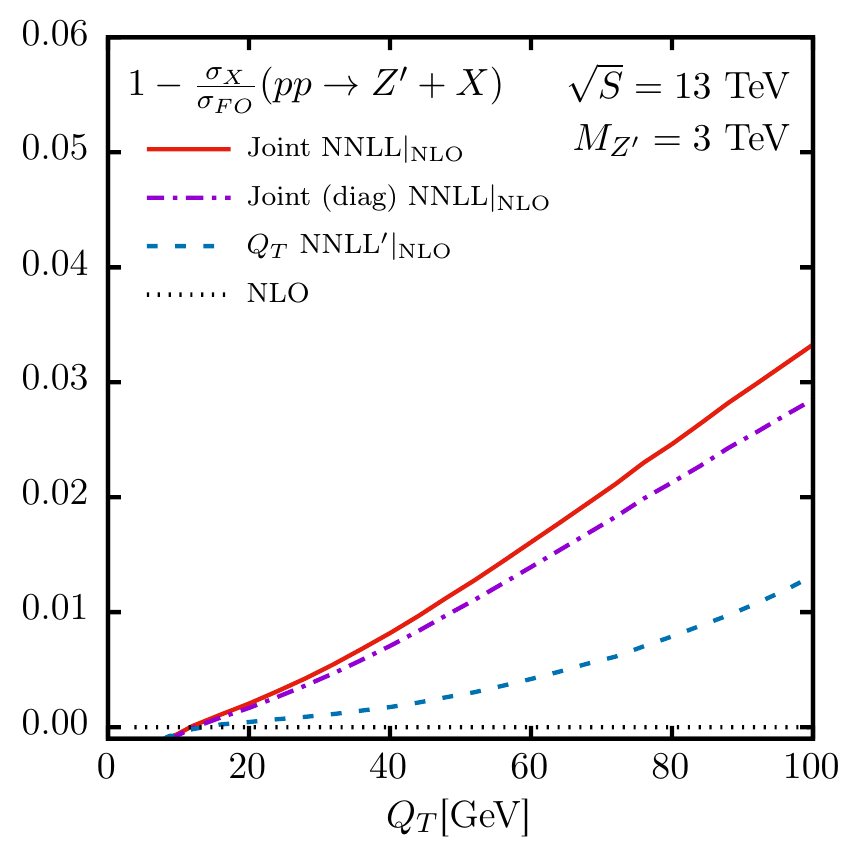} &
		\includegraphics[width=0.45\textwidth]{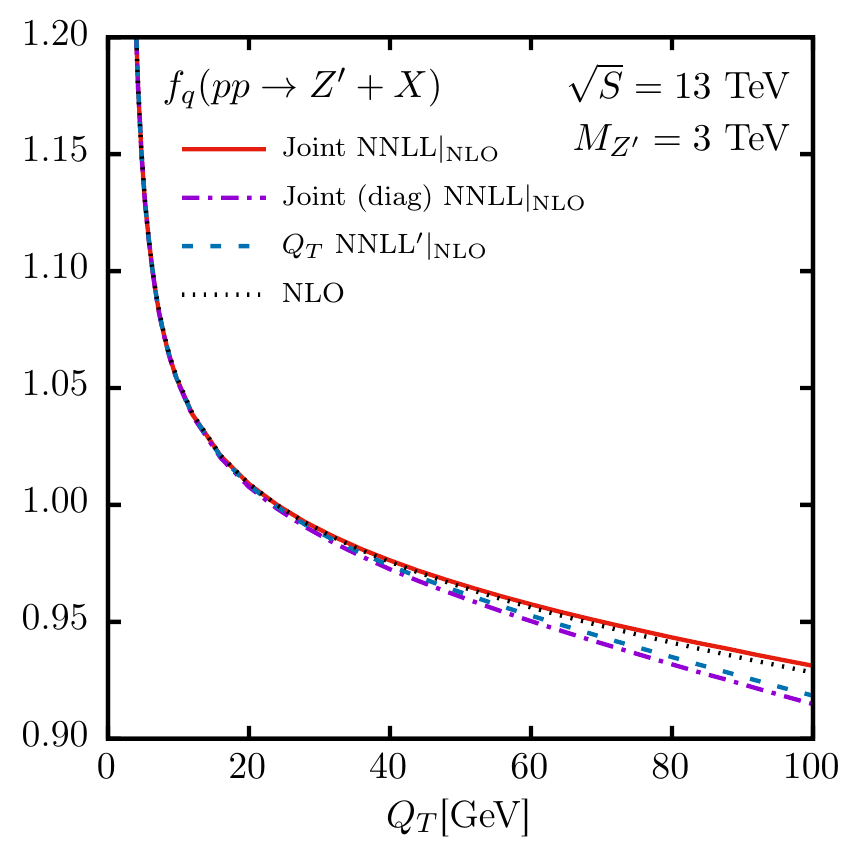} \tabularnewline
		\hspace{1.7em} (a) &\hspace{1.7em} (b) \tabularnewline
	\end{tabular}
	\caption{The same as Fig.~\ref{fig:expansion-LHC-Zp-NLL},  but comparing the expansion of the NNLL resummation to NLO accuracy.} 
	\label{fig:expansion-LHC-Zp-NNLL}
\end{figure}

In order to analyze a process closer to threshold we study $Z'$ production. A mass $M_{Z'}=3$ TeV is used and the other parameters are kept the same as for $Z$-boson production. In order to improve the fit of the PDFs in Mellin space at these scales, the same implementation as in the code {\sc Resummino}~\cite{Fuks:2013vua} was used.
As can be seen from Fig.~\ref{fig:expansion-LHC-Zp-NLL}(a) all three expansions provide a good approximation of the fixed order. The size of threshold effects which are not already captured by the $Q_T$ formalism is rather small at central scale.
In addition, it can be noted that the two different methods of joint resummation now agree. The reason for this can be seen in Fig.~\ref{fig:expansion-LHC-Zp-NLL}(b). For $Z'$ production with a high enough mass the dominant channel is the $q\bar{q}$ and therefore the difference between the two methods of joint resummation will be small.
Finally the $Z'$ $Q_T$-distribution at NLO accuracy can be seen in Fig.~\ref{fig:expansion-LHC-Zp-NNLL}. Fig.~\ref{fig:expansion-LHC-Zp-NNLL}(a) shows that the expansion is slightly worse for joint resummation when compared to $Q_T$ resummation, however this difference is around the 1\% level. The $q\bar{q}$ fraction, as seen in Figure~\ref{fig:expansion-LHC-Zp-NNLL}, is slightly better for joint resummation.

\subsection{Resummation}

Having explored the regime of validity of the expansion, we now focus our attention on the effect of joint resummation on the transverse momentum distribution and its theoretical uncertainties. 

\begin{figure}
	\centering
	\begin{tabular}{cc}
		\includegraphics[width=0.45\textwidth]{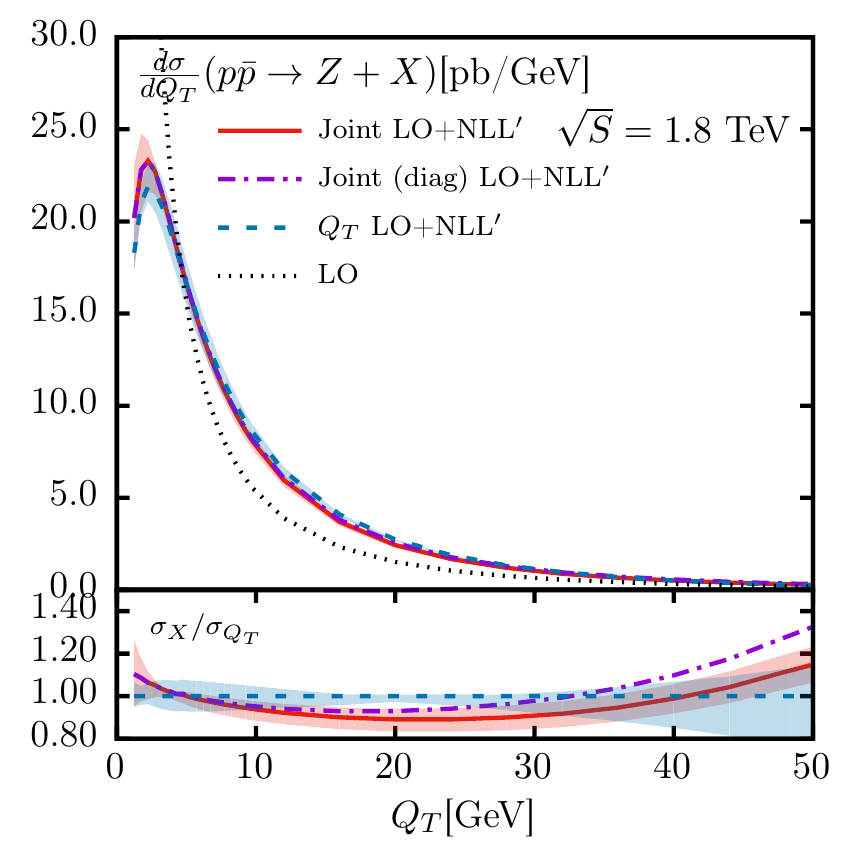} &
		\includegraphics[width=0.45\textwidth]{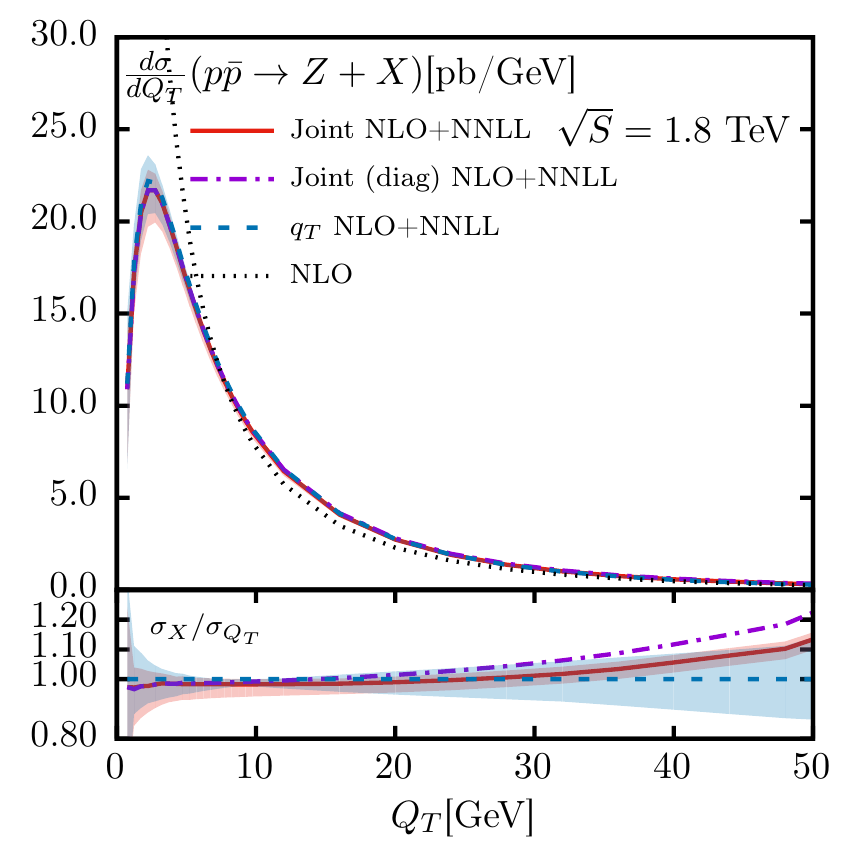} \tabularnewline
		\hspace{1.7em} (a) &\hspace{1.7em} (b) \tabularnewline
	\end{tabular}
	\caption{The $Z$-boson transverse momentum distribution at $\sqrt{S}=1.8$ TeV Tevatron collision energy. Fixed-order and resummed and matched results are compared at different perturbative accuracies.  The uncertainty bands are computed by independently varying the factorization and renormalization scales with the 7-point method. The lower panel shows the ratio with respect to the central value of $Q_T$ resummation.} 
	\label{fig:resummation-Teva}
\end{figure}

We begin by showing resummed results in the Tevatron setup. In Fig.~\ref{fig:resummation-Teva} the transverse momentum distribution is shown in fixed-order perturbation theory and resummed perturbation theory. In particular, the plot in Fig.~\ref{fig:resummation-Teva}(a) shows LO (dotted) and LO+NLL$^\prime$ for joint (solid), joint diagonal (dotted-dashed) and $Q_T$ resummation. 
Uncertainty bands are provided for $Q_T$ resummation and joint resummation and they are determined by independently varying the factorization and renormalization scale by a factor 2 using the 7-point method. 
The bottom panel shows the ratio of the joint-resummation result to standard $Q_T$ resummation at the central scale.
Joint resummation causes a small increase in the peak region, followed by a decrease for larger values of $Q_T$ and finally an increase in the tail. In the low $Q_T$ region the two methods of joint resummation agree and for increasing $Q_T$ values the difference becomes larger. This shows that it is important to have a correct representation of the power corrections in order to have good control at larger values of $Q_T$.
We note that, with the exception of the region roughly between 15 and 30 GeV, joint resummation does reduce the uncertainty. We believe that scale variation in $Q_T$ resummation underestimates the uncertainty because the curves for different scales have a pinch point in this region, while the pinch for joint resummation is less pronounced and it appears at lower transverse momentum, in the $Q_T \sim 5$~GeV region where $Q_T$ resummation is dominant.

Next the we consider the $Q_T$ spectrum one order higher in perturbation theory.
In Fig.~\ref{fig:resummation-Teva}(b) we plot NLO (dotted), NLO+NNLL for joint (solid) and joint diagonal (dotted-dashed), and NLO+NNLL$^\prime$ for $Q_T$ resummation.
As expected, joint resummation further reduces the scale uncertainty. 
In addition the difference between the two methods of joint is smaller at this accuracy, albeit outside the uncertainty band. 
The behavior now also changes in comparison to LO+NLL$^\prime$ accuracy. In the low $Q_T$ region, joint resummation agrees with $Q_T$ resummation, while there is still an increase in the tail region. We believe this to be an indication that $Q_T$ resummation alone, if considered at high-enough orders, does capture most of the threshold effects. 

\begin{figure}
	\centering
	\begin{tabular}{cc}
		\includegraphics[width=0.45\textwidth]{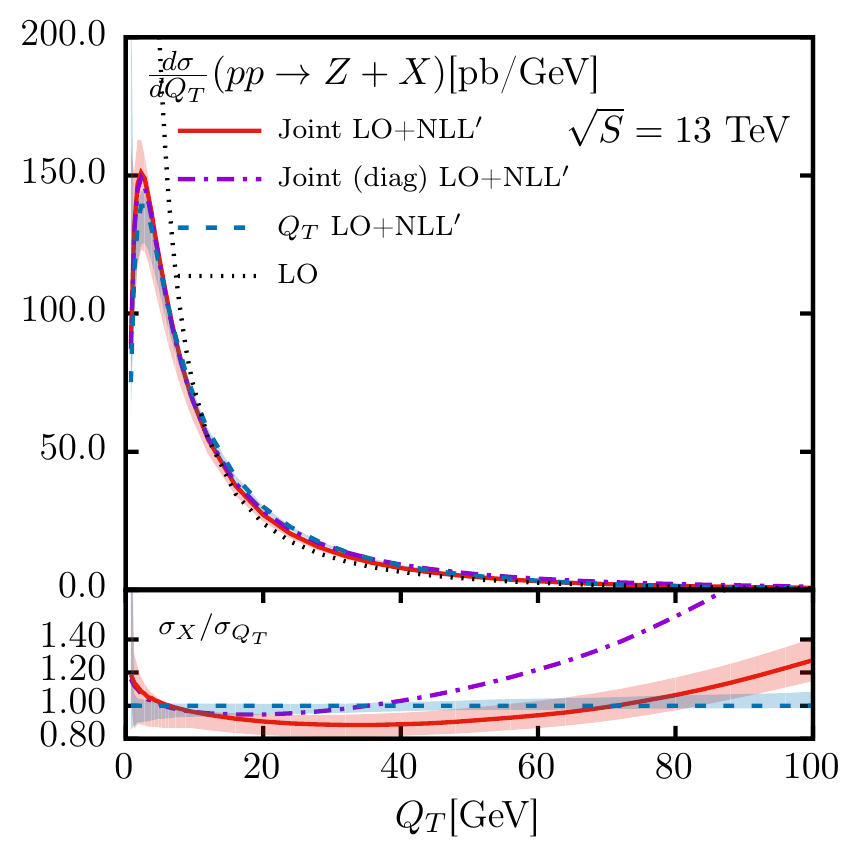} &
		\includegraphics[width=0.45\textwidth]{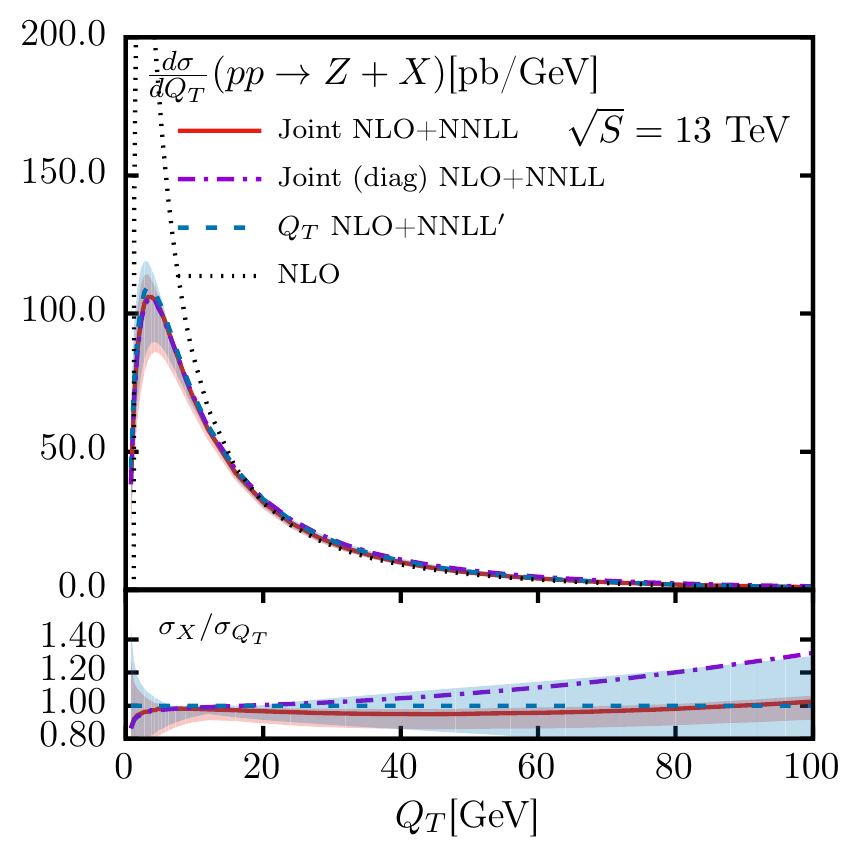} \tabularnewline
		\hspace{1.7em} (a) &\hspace{1.7em} (b) \tabularnewline
	\end{tabular}
	\caption{The same as Figure~\ref{fig:resummation-Teva}, for the LHC at $\sqrt{S}=13$ TeV.} 
	\label{fig:resummation-LHC}
\end{figure}

We have already seen that the expansion of joint resummation does not approximate fixed order any better than $Q_T$ resummation in the case of $Z$ production at the LHC. However, it is still interesting to look at the behavior the resummed cross section would have. This result is presented in Fig.~\ref{fig:resummation-LHC}. We note that the behavior of joint resummation at LO+NLL$^\prime$ with respect to $Q_T$ resummation is comparable to the lower-energy (Tevatron) case.
However,  we do notice a significant difference between the two methods of joint resummation, which indicates a strong dependence on the power corrections. This difference becomes smaller at NLO+NNLL accuracy, but it is still significant.

\begin{figure} 
	\centering
	\begin{tabular}{cc}
		\includegraphics[width=0.45\textwidth]{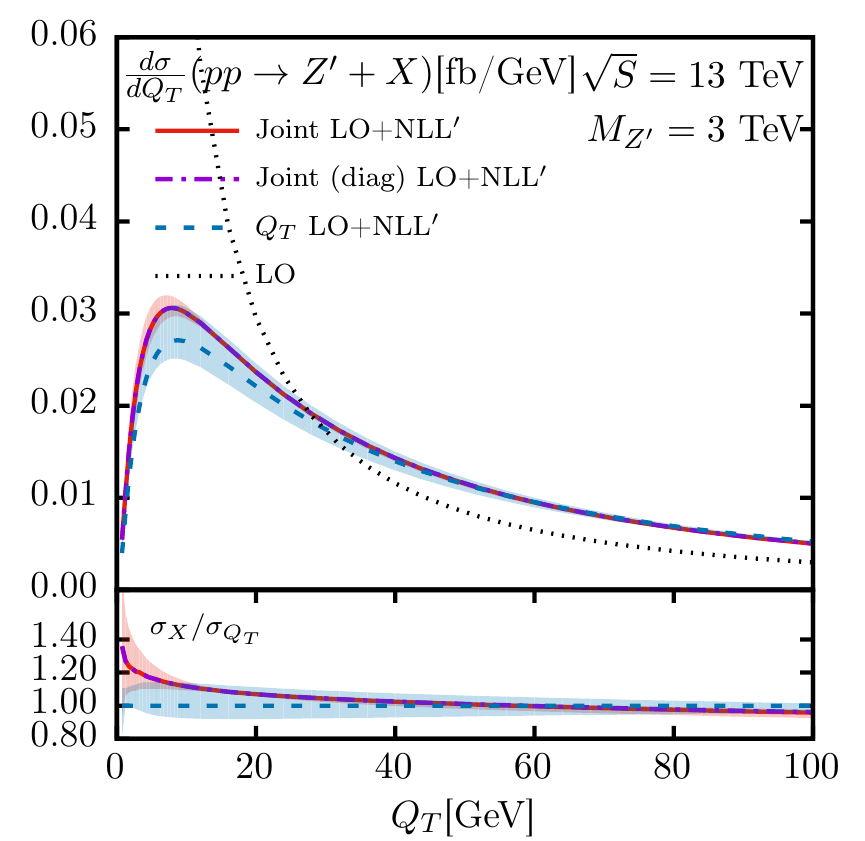} &
		\includegraphics[width=0.45\textwidth]{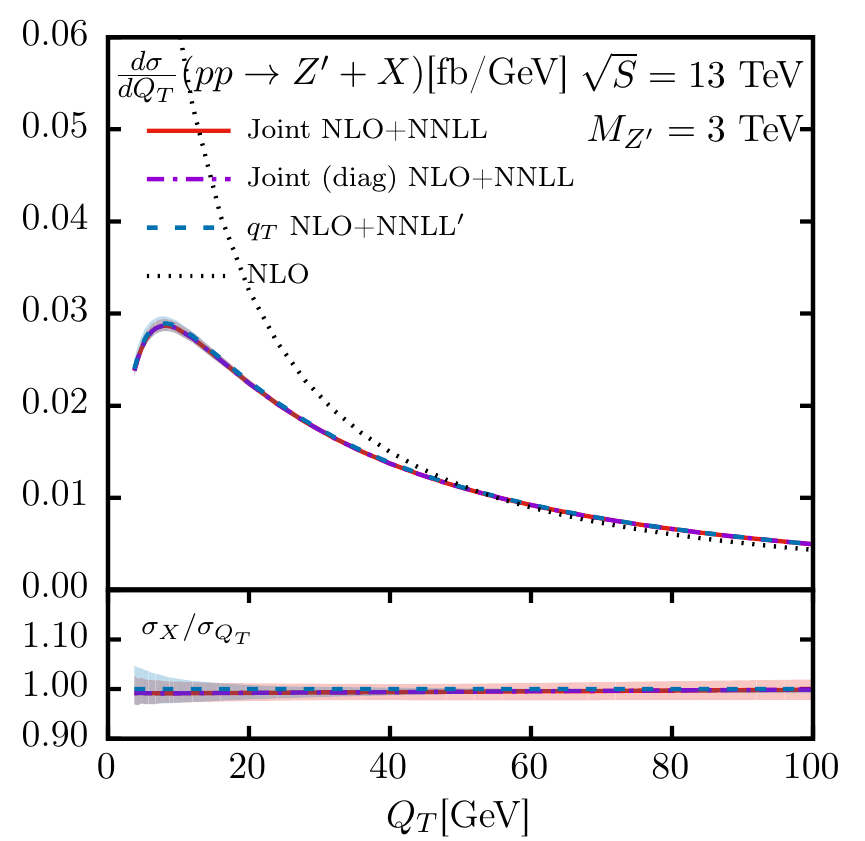} \tabularnewline
		\hspace{1.7em} (a) &\hspace{1.7em} (b) \tabularnewline
	\end{tabular}
	\caption{The same as Figure~\ref{fig:resummation-LHC}, but for $Z'$ production with $M_{Z'}=3$ TeV.} 
	\label{fig:resummation-LHC-Zp}
\end{figure}

Finally, in Fig.~\ref{fig:resummation-LHC-Zp}, we show resummed results for $Z'$ production . At LO+NLL$^\prime$ accuracy, shown in (a), an increase can be seen for the low values of $Q_T$. The two different methods for joint resummation do agree with each another. In addition, a significant reduction of the scale dependence can be observed. At NLO+NNLL accuracy, shown in (b), joint resummation and $Q_T$ resummation provide very similar result but we do notice a further decrease in the scale uncertainty.

\section{Conclusions}\label{sec:conclusion}
In this paper we have considered the transverse momentum distribution of an electro-weak vector boson in joint resummation. This formalism, which was first developed in Refs.~\cite{Li:1998is,Laenen:2000ij} allows for the simultaneous resummation of logarithmic contributions that are enhanced at small $Q_T$ and those that are enhanced at threshold.
While phenomenological applications of this formalism existed at NLL~\cite{Kulesza:2002rh,Kulesza:2003wn,Banfi:2004xa}, to our knowledge, no analysis was performed beyond this logarithmic accuracy.  
In this paper, we have derived and implemented joint resummation at NNLL. In particular, we have considered the production of a $Z$ boson via the DY mechanism at the Tevatron and at the LHC, as well as of a heavier $Z^\prime$ at the LHC. 
By comparing fixed-order results with their approximations obtained by expanding the joint-resummed result in powers of the strong coupling, we have performed a detailed study of the regime of validity for our implementation. 
We have found that its use is fully justified for $Z$ production at the Tevatron, while at the LHC the situation is much less clear, because there are significant contributions from power-corrections to the threshold limit. For instance, the $q\bar{q}$ channel is not the dominant channel away from the small $Q_T$ limit.
On the other hand, the formalism works well if the production of heavier particles, such as a hypothetical $Z'$ with a mass of 3~TeV, is considered. 
When looking at all-order results, we have found that joint resummation at NLL$^\prime$ gives noticeable corrections when compared to standard $Q_T$ resummation at the same accuracy. However, differences between the two are much smaller when both resummations are upgraded to NNLL. Nevertheless, NNLL joint resummation leads to a further decrease of the scale dependence. 

We see several possible directions for future developments of this work. The first one, more theoretical, consists of revisiting the original derivation of Ref.~\cite{Laenen:2000ij} in order to better understand the role of power corrections to the threshold limit with the aim of including them in phenomenological studies. This will put the joint resummation of Standard Model processes at LHC energies on a firmer ground. 
Moreover, it would be interesting to quantitatively compare the approach presented in this work to the threshold resummation of the $Q_T$ spectrum, as done for instance in Refs.~\cite{Gonsalves:2005ng,Kidonakis:2014zva,deFlorian:2005fzc}.
A special, and particularly interesting, case is given by Higgs production in gluon gluon fusion, in which power-suppressed contributions at threshold are known to play a less important role than in DY. 
In this context, we plan  to explore the possibility of combining these results with other kinds of joint resummation, such as the simultaneous resummation of small- and large-$x$~\cite{Ball:2013bra} contributions, as well the joint resummation of small-$x$ and $Q_T$ logarithms, recently proposed in Ref.~\cite{Marzani:2015oyb}.
Furthermore, one could also concentrate on Beyond the Standard Model processes. For instance, one could imagine to apply our results to the production of supersymmetric particles and therefore upgrade the accuracy of the computer code {\sc Resummino}~\cite{Fuks:2013vua} to NNLL. 

\begin{acknowledgments}
We thank Giancarlo Ferrera, Anna Kulesza, and Eric Laenen for useful discussions and Claudio Muselli for a critical reading of the manuscript.
This work was supported by the U.S.\ National Science Foundation, under grants PHY-0969510, the LHC Theory Initiative, PHY-1417317 and PHY-1619867.
Support was provided by the Center for Computational Research at the University at Buffalo.
\end{acknowledgments}

\appendix
\section{Resummation coefficients}\label{sec:coef}
In this section we list the coefficients that enters our joint resummation formula Eq.~(\ref{eq:joint-exponent-final}), focusing on the flavor-diagonal $q\bar q$ contributions. We start with the first three coefficient of the cusp
\begin{align}
A^{(1)}&=C_F\,,\\
A^{(2)}&=\frac{1}{2}C_F\left[C_A\left(\frac{67}{18}-\frac{\pi^2}{6}\right)-\frac{5}{9}n_f\right]\,, \\
A^{(3)}&=\frac{1}{4}C_F\left[C_A^2\left(\frac{245}{24}-\frac{67\pi^2}{54}+\frac{11}{6}\zeta_3+\frac{11\pi^4}{180}\right)+C_Fn_f\left(-\frac{55}{24}+2\zeta_3\right)\right.\nonumber\\
&\left.+C_An_f\left(-\frac{209}{108}+\frac{5\pi^2}{27}-\frac{7}{3}\zeta_3\right)-\frac{n_f^2}{27}\right].
\end{align}
The soft-wide angle contribution is given by:
\begin{align}
\tilde{D}^{(1)}&=0\,,\\ \tilde{D}^{(2)}&=C_F\left[C_A\left(-\frac{101}{27}+\frac{7}{2}\zeta_3\right)+\frac{14}{27}n_f\right].
\end{align}
Note that $A^{(3)}_\text{thr}=A^{(3)}$, and $A^{(3)}_{Q_T}=A^{(3)}_\text{thr}- \beta_0 \tilde{D}^{(2)}$.
Furthermore, we can write
\begin{equation}
\tilde{B}(N, b,\as)= \sum_{k=1}^\infty \left( \frac{\as}{\pi}\right)^k B^{(k)}(N,b),
\end{equation}
with
\begin{align}
\tilde{B}^{(1)}(N,b)&=-\frac{3}{2}C_F+2\gamma^{(0)}(N)=C_F\left(\frac{1}{N(N+1)}-2\left(\psi(N+1)+\gammae\right)\right),\\
\tilde{B}^{(2)}(N,b)&=C_F\left[C_F\left(\frac{\pi^2}{4}-\frac{3}{16}-3\zeta_3\right)+C_A\left(-\frac{11\pi^2}{72}-\frac{17}{48}+\frac{3}{2}\zeta_3\right)+n_f\left(\frac{1}{24}+\frac{\pi^2}{24}\right)\right]\nonumber\\
&+2\gamma^{(1)}(N)-\beta_0 C_F\left[\zeta_{2}-{\rm Li}_{2}\left(\frac{\bar{b}^{2}}{\chi^2}\right)\right]-2\beta_0\frac{C_F}{N(N+1)},
\end{align}
where $\gamma^{(0)}(N)$ and $\gamma^{(1)}(N)$ are the one- and two-loop $qq$ DGLAP anomalous dimension, respectively.
Note that, in the threshold limit $N \to \infty$, the above coefficients reduce to
\begin{align}
\tilde{B}^{(1)}&=-2 A^{(1)} \log\bar{N}+\Ord(1/N),\\
\tilde{B}^{(2)}&=-2 A^{(2)} \log\bar{N}+\Ord(1/N),
\end{align}
with no constant contribution, as expected.
Next, we move to the one-loop hard coefficient in joint resummation:
\begin{align}\label{Hfactexplicit}
\mathcal{H}^{F,(1)}_{\text{joint}}(N,b,\muf)&=C_F\left(3\zeta_2-4+\frac{1}{N(N+1)}\right)+C_F\left[\zeta_{2}-{\rm Li}_{2}\left(\frac{\bar{b}^{2}}{\chi^2}\right)\right]\nonumber\\&+2\left(\gamma^{(0)}_{qq}(N)+C_F\log\bar{N}\right)\log\left(\frac{Q^2}{\muf^2}\right).
\end{align}
Note that the additional $\log\bar{N}$ in Eq.~(\ref{Hfactexplicit}) subtracts off the large-$N$ behavior of the anomalous dimensions, which is resummed in the exponent. In the large-$N$ limit we recover $\mathcal{C}^{(1)}=C_F\left[4\zeta_2-4\right]$, while at large $b$ we obtain the standard $Q_T$-resummation result $\mathcal{H}^{F,(1)}(N)=C_F\left[3\zeta_2-4+1/(N(N+1))\right]$, which are both given at $\muf=Q$.
Finally, at NNLL, the two-loop hard coefficient is the same as for $Q_T$ resummation and will not be listed here. 

Next, we detail the necessary steps to derive Eq.~(\ref{eq:joint-exponent-final}) from Eq.~(\ref{eq:joint-exponent}). The main focus will be rewriting the term that contributes to the difference in $A^{(3)}_{Q_T}$ and  $A^{(3)}_{\text{thr}}$. For this we look at the sum of the wide-angle and cusp  contributions:
\begin{align}\label{eq:col-anom+WA}
\Delta \Gj^{\mathrm{cusp}}+\Delta \Gj^\mathrm{wide-angle}&=\int_{Q^2/\chi^2}^{Q^2/\bar{N}^2}\frac{dq^2}{q^2}\beta_0\tilde{D}^{(2)}\left(\frac{\as(q)}{\pi}\right)^{3}\log\frac{Q^2}{q^2}-\frac{1}{2}\int^{Q^2}_{Q^2/\bar{N}^2}\frac{dq^2}{q^2}\tilde{D}(\as(q))\\
&=-\frac{1}{2}\int_{Q^2/\chi^2}^{Q^2/\bar{N}^2}\frac{dq^2}{q^2}\beta(\as(q))\frac{\partial\tilde{D}(\as(q))}{\partial\as}\log\frac{Q^2}{q^2}-\frac{1}{2}\int^{Q^2}_{Q^2/\bar{N}^2}\frac{dq^2}{q^2}\tilde{D}(\as(q)),\nonumber
\end{align}
where the second line in the above equation holds up to NNLL accuracy.
We rewrite Eq.~(\ref{eq:col-anom+WA}) as 
\begin{align}\label{eq:col-anom+WA2}
\Delta \Gj^{\mathrm{cusp}}+\Delta \Gj^\mathrm{wide-angle}&=-\frac{1}{2}\int_{Q^2/\chi^2}^{Q^2/\bar{N}^2}\frac{dq^2}{q^2}\int_{q^2}^{Q^2}\frac{dk^2}{k^2}\beta(\as(q))\frac{\partial\tilde{D}(\as(q))}{\partial\as}-\frac{1}{2}\int^{Q^2}_{Q^2/\bar{N}^2}\frac{dq^2}{q^2}\tilde{D}(\as(q))\nonumber\\
&=-\frac{1}{2}\int_{Q^2/\chi^2}^{Q^2/\bar{N}^2}\frac{dk^2}{k^2}\int_{Q^2/\chi^2}^{k^2}\frac{dq^2}{q^2}\beta(\as(q))\frac{\partial\tilde{D}(\as(q))}{\partial\as}-\frac{1}{2}\int^{Q^2}_{Q^2/\bar{N}^2}\frac{dq^2}{q^2}\tilde{D}(\as(q))\nonumber\\
&-\frac{1}{2}\int_{Q^2/\bar{N}^2}^{Q^2}\frac{dk^2}{k^2}\int_{Q^2/\chi^2}^{Q^2/\bar{N}^2}\frac{dq^2}{q^2}\beta(\as(q))\frac{\partial\tilde{D}(\as(q))}{\partial\as}.
\end{align}
Furthermore, using the identity
\begin{equation}\label{eq:running-D}
\int_{q_2^2}^{q_1^2}\frac{dq^2}{q^2}\beta(\as(q)) \frac{\partial\tilde{D}(\as(q))}{\partial\as} = \tilde{D}(\as(q_1))-\tilde{D}(\as(q_2)),
\end{equation}
we obtain
\begin{align}\label{eq:col-anom+WA3}
\Delta \Gj^{\mathrm{cusp}}+\Delta \Gj^\mathrm{wide-angle}&=-\frac{1}{2}\int_{Q^2/\chi^2}^{Q^2/\bar{N}^2}\frac{dk^2}{k^2}\left(\tilde{D}(\as(k))-\tilde{D}(\as(Q/\chi))\right)-\frac{1}{2}\int^{Q^2}_{Q^2/\bar{N}^2}\frac{dq^2}{q^2}\tilde{D}(\as(q))\nonumber\\
&-\frac{1}{2}\int_{Q^2/\bar{N}^2}^{Q^2}\frac{dk^2}{k^2}\left(\tilde{D}(\as(Q/\bar{N}))-\tilde{D}(\as(Q/\chi))\right)\nonumber\\
&=-\frac{1}{2}\int^{Q^2}_{Q^2/\chi^2}\frac{dq^2}{q^2}\tilde{D}(\as(q))-\frac{1}{2}\left(\tilde{D}(\as(Q/\bar{N}))\log\bar{N}^2-\tilde{D}(\as(Q/\chi))\log\chi^2\right)\nonumber\\
&=-\frac{1}{2}\int^{Q^2}_{Q^2/\chi^2}\frac{dq^2}{q^2}\tilde{D}(\as(q))-\frac{1}{2}\tilde{D}(\as(Q/\chi))\log\frac{\bar{N}^2}{\chi^2},
\end{align}
where the difference in scales in the final step is beyond NNLL accuracy in both $Q_T$ and threshold resummation.
Finally, we verify that in the threshold limit  Eq.~(\ref{eq:joint-exponent-final}) reduces to
\begin{align}\label{eq:joint-exponent-threshold}
\Gj^{\mathrm{NNLL}}(\as(\mur),b=0,N,Q^2/\mur^2)=& -\int_{Q^2/\bar{N}^2}^{Q^2}\frac{dq^2}{q^2}\Bigg[A(\as(q))\log\frac{Q^2}{q^2}+\tilde{B}(N,b=0,\as(q))+\frac{1}{2}\tilde{D}(\as(q))\Bigg]\nonumber \\&+2\int_{\muf^2}^{Q^2}\frac{dq^2}{q^2}\gamma_\text{soft}(N,\as(q))\nonumber\\
=&-\int_{Q^2/\bar{N}^2}^{Q^2}\frac{dq^2}{q^2}\Bigg[A(\as(q))\log\frac{Q^2}{q^2}-2A(\as(q))\log\bar{N}+\frac{1}{2}\tilde{D}(\as(q))\Bigg]\nonumber \\&+2\int_{\muf^2}^{Q^2}\frac{dq^2}{q^2}\gamma_\text{soft}(N,\as(q))+\Ord(1/N),
\end{align}
which agrees with Eq.~(\ref{eq:sudakov-thr}).
For the $Q_T$ resummation limit it is easier to look at the exponent before rewriting the difference between the $A^{(3)}$ contributions. This expression in the limit $\chi\to\bar{b}$ results in: 
\begin{align}\label{eq:joint-exponent-QT}
\Gj^{\mathrm{NNLL}}(\as(\mur),b,N,Q^2/\mur^2)\underset{\chi\to\bar{b}}{=}& -\int_{Q^2/\bar{b}^2}^{Q^2}\frac{dq^2}{q^2}\Bigg[A_{Q_T}(\as(q))\log\frac{Q^2}{q^2}+\tilde{B}(b,\as(q))\Bigg]\nonumber \\&+\frac{1}{2}\int^{Q^2}_{Q^2/\bar{N}^2}\frac{dq^2}{q^2}\beta(\as(q))\frac{\partial\tilde{D}(\as(q))}{\partial\as}\log\frac{Q^2}{q^2}+2\int_{\muf^2}^{Q^2}\frac{dq^2}{q^2}\gamma_\text{soft}(N,\as(q))\nonumber.
\end{align}
This contribution agrees with Eq.~(\ref{eq:sudakov-qt}) up to the additional exponentiation of constant terms.

\section{Approximation of Fourier transform for logarithms}\label{sec:J0}
In order to simplify the Fourier transform we will approximate the Bessel function $J_{0}$. We follow the approach of Appendix A of \cite{Catani:2003zt} with the necessary changes in order to address the specific case of $Q_{T}$ resummation.
We are interested in evaluating integrals of the form
\begin{equation}
I_{n}\left(b\right)\equiv\int_{0}^{1}\frac{dx}{x}\left(1-J_{0}\left(\tilde{b}x\right)\right)\log^{n}x,
\end{equation}
where $x$ is a dimensionless version of $k_{T}$ ($x=k_{T}/Q$) and $\tilde{b}$ is a dimensionless version of $b$ ($\tilde{b}=Qb$). In order to perform the integral we use a generating function:
\begin{equation}
\log^{n}x=\lim_{\epsilon\to0}\left(\frac{\partial}{\partial\epsilon}\right)^{n}x^{\epsilon},
\end{equation}
which leads to 
\begin{eqnarray}
I_{n}\left(b\right) & = & \lim_{\epsilon\to0}\left(\frac{\partial}{\partial\epsilon}\right)^{n}\left\{ \frac{1}{\epsilon}\left[1-{}_{1}F_{2}\left(\frac{\epsilon}{2};1,1+\frac{\epsilon}{2};-\frac{\tilde{b}^{2}}{4}\right)\right]\right\} \nonumber \\
&= & \lim_{\epsilon\to0}\left(\frac{\partial}{\partial\epsilon}\right)^{n}\left\{ \frac{1}{\epsilon}\left[1-\left(\frac{2}{\tilde{b}}\right)^{\epsilon}\frac{\Gamma\left(1+\frac{\epsilon}{2}\right)}{\Gamma\left(1-\frac{\epsilon}{2}\right)}+\Ord\left( \frac{1}{\tilde{b}}\right)\right]\right\},
\end{eqnarray}
where we have approximated in the large $\tilde{b}$ limit. This can be rewritten as
\begin{equation}\label{eq:In}
I_{n}\left(b\right)=\lim_{\epsilon\to0}\left(\frac{\partial}{\partial\epsilon}\right)^{n}\left\{ \frac{1}{\epsilon}\left[1-\exp\left[-\epsilon\left(\log\left(\tilde{b}/2\right)+\gamma_{E}\right)+\sum_{n=1}^{\infty}\zeta_{2n+1}\frac{\epsilon^{2n+1}}{2^{2n}\left(2n+1\right)}\right]+\Ord\left( \frac{1}{\tilde{b}}\right)\right]\right\},
\end{equation}
with the Riemann zeta function $\zeta_{i}$. Now a function $\Psi$ can be defined as
\begin{equation}
e^{-\epsilon\log\bar{b}}\exp\left[\sum_{n=1}^{\infty}\zeta_{2n+1}\frac{\epsilon^{2n+1}}{2^{2n}\left(2n+1\right)}\right]=\Psi\left(1-\frac{\partial}{\partial\log\bar{b}}\right)e^{-\epsilon\log\bar{b}},
\end{equation}
with $\bar{b}=\tilde{b}e^{\gamma_{E}}/2=bQe^{\gamma_{E}}/2$ and
\begin{equation}\label{eq:Psi}
\Psi\left(1+\epsilon\right)\equiv\exp\left[\sum_{n=1}^{\infty}\zeta_{2n+1}\frac{\epsilon^{2n+1}}{2^{2n}\left(2n+1\right)}\right],
\end{equation}
By taking the nth order derivative with respect to $\epsilon$ after filling it back into equation~(\ref{eq:In}) we obtain:
\begin{equation}
I_{n}\left(b\right)=\Psi\left(1-\frac{\partial}{\partial\log\bar{b}}\right)\frac{\left(-\log\bar{b}\right)^{n+1}}{n+1}+{\cal O}\left(\frac{1}{b}\right).
\end{equation}
Finally by noting that
\begin{equation}
\frac{\left(-\log b\right)^{n+1}}{n+1}=-\int_{1/b}^{1}dx\frac{\log^{n}x}{x},
\end{equation}
we can approximate 
\begin{equation}
1-J_{0}\left(bk_{T}\right)=\Psi\left(1-\frac{\partial}{\partial\log\bar{b}}\right)\Theta\left(k_{T}-Q/\bar{b}\right)+{\cal O}\left(\frac{1}{b}\right),
\end{equation}
The Riemann zeta functions from $\Psi$ only start contributing at N$^{3}$LL order, because the series in $\epsilon$ in equation~(\ref{eq:Psi}) only starts at $\epsilon^{3}$. Therefore at NNLL accuracy we can use the approximation:
\begin{equation}
1-J_{0}\left(bk_{T}\right)\approx\Theta\left(k_{T}-Q/\bar{b}\right).
\end{equation}

\section{The joint resummation function $\chi$}\label{sec:chi}
There are two main points to take into account when choosing the function $\chi$. Firstly the choice of $\chi$ influences the power suppressed terms that result from the expansion of $\log(\chi)$. This can be seen by expanding this logarithm in either the large $\bar{b}$ or large $\bar{N}$ limit. For example we can look at the choice $\chi=\bar{b}+\bar{N}$:
\begin{align}
\log(\chi)&\underset{\bar{b}\to\infty}{=}\log(\bar{b})+\frac{\bar{N}}{\bar{b}}-\frac{\bar{N}^2}{2\bar{b}^2}+\mathcal{O}(\bar{b}^{-3}),\\
\log(\chi)&\underset{\bar{N}\to\infty}{=}\log(\bar{N})+\frac{\bar{b}}{\bar{N}}-\frac{\bar{b}^2}{2\bar{N}^2}+\mathcal{O}(\bar{N}^{-3}).\\
\end{align}
This introduces $\bar{b}^{-1}$ power corrections in $Q_T$ resummation, which are not present in fixed-order calculation. Note that the power suppressed terms in threshold resummation are not as important, because they will not contribute if $Q_T$ is integrated over.
Alternatively $\chi=\bar{b}+\frac{\bar{N}^{2}}{\bar{N}+\bar{b}\eta}$ can be defined~\cite{Kulesza:2003wn}. In this case the power suppressed contribution in the $Q_T$ resummation limit starts at $\bar{b}^{-2}$:
\begin{align}
\log(\chi)&\underset{\bar{b}\to\infty}{=}\log(\bar{b})+\frac{\bar{N}^2}{\eta\bar{b}^2}+\mathcal{O}(\bar{b}^{-3}),\\
\log(\chi)&\underset{\bar{N}\to\infty}{=}\log(\bar{N})+\frac{\bar{b}}{\bar{N}}(1-\eta)-\frac{\bar{b}^2}{2\bar{N}^2}(1-2\eta-\eta^2)+\mathcal{O}(\bar{N}^{-3}).
\end{align}
Because of the behavior of the power corrections, this is the preferred choice of $\chi$.

The second feature that differs between between possible choices of $\chi$ is the choice of the contour to be used to compute the Fourier and Mellin inverse transformation. Particular attention must be paid to the singularity structure of the integrand. Unlike for separate threshold and $Q_T$ resummation, in joint resummation the Mellin space variable, $N$, and the Fourier space variable, $b$, are connected through one function $\chi$. The singularities are points in the $\chi$-plane, therefore in the $b$-plane these are lines, which depend on the choice for the function $\chi$. The potential singularities are at the values 0, $\infty$ and the Landau pole $\rho_L=\exp[1/(2b_0\as)]$ for $\chi$. 
If the Fourier and Mellin variables are parameterized as:
\begin{align}
b&=x_b e^{\pm I\phi_b},\nonumber\\
N&=c+x_N e^{\pm I\phi_N}
\end{align}
we can relate the choice of angles of the different variables to one another.

\begin{figure} 
	\centering
	\begin{tabular}{c}
		\includegraphics[width=0.45\textwidth]{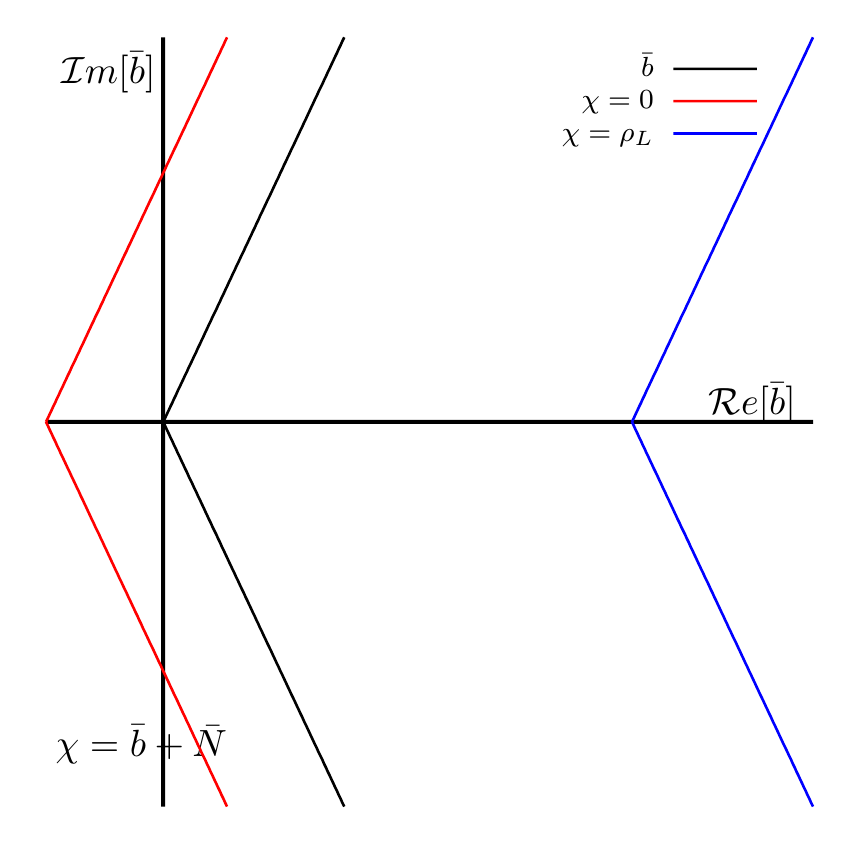} \tabularnewline
		\hspace{1.7em} (a)  \tabularnewline
	\end{tabular}

	\begin{tabular}{cc}
		\includegraphics[width=0.45\textwidth]{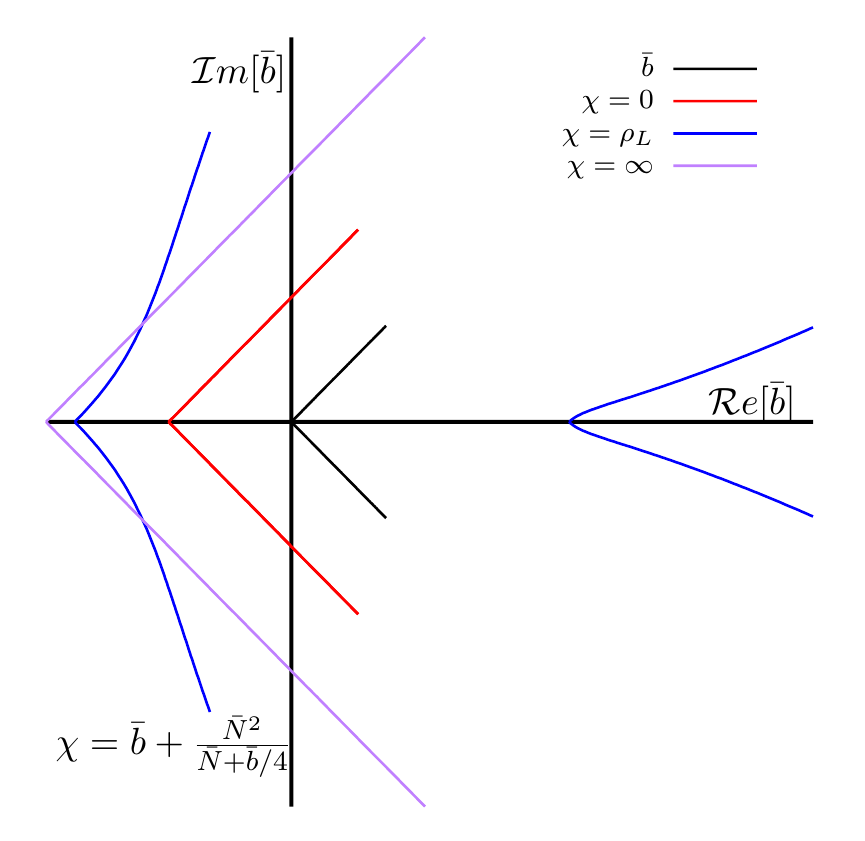} &
		\includegraphics[width=0.45\textwidth]{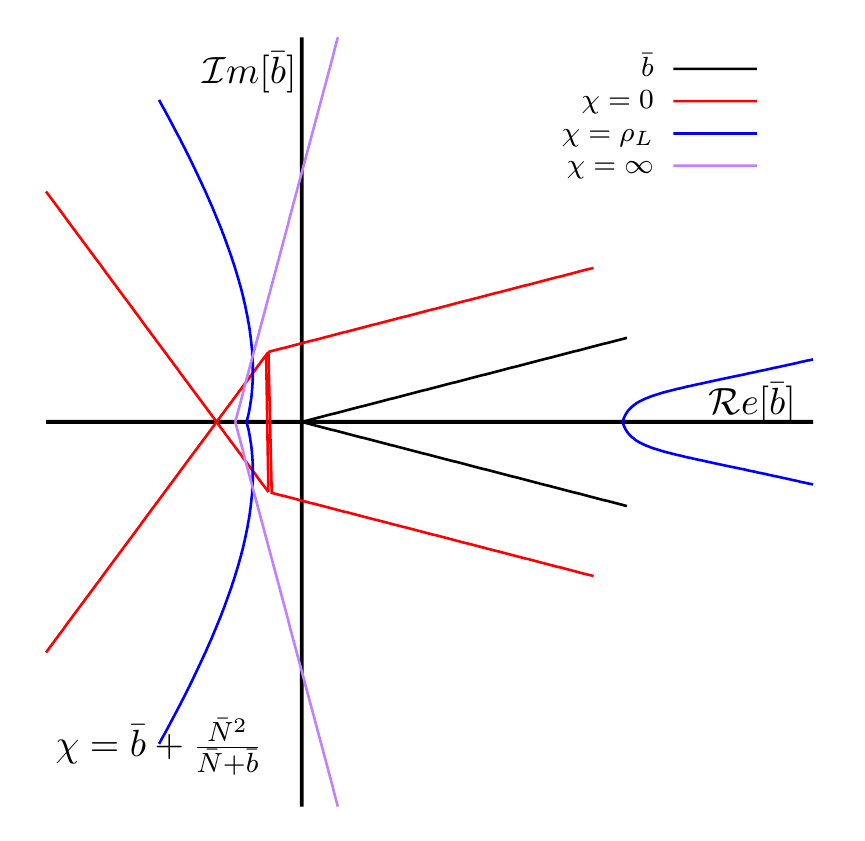} \tabularnewline
		\hspace{1.7em} (b) &\hspace{1.7em} (c) \tabularnewline
	\end{tabular}
	\caption{The $\bar{b}$ contour and the different solutions to the $\chi$ singularities plotted in terms of the real and imaginary parts of $\bar{b}$. In black the $\bar{b}$ contour is shown for a variation of the integration parameter $x_b$ and in red, blue and purple the different singularities as a function of the $\bar{N}$ contour through variation of $x_N$. This is shown for different choices of $\chi$ in (a), (b) and (c).}
	\label{fig:contour-chi}
\end{figure}

First we can look at the simplest choice: $\chi=\bar{b}+\bar{N}$. Here the solutions for the singularities are straight lines in the complex plain and are listed in Table~\ref{tab:poles}. In order to avoid these lines the angle $\phi_b$ should be chosen so that the $b$-contour line is parallel to $-\bar{N}$, as can be seen in Fig,~\ref{fig:contour-chi}(a). This leads to the requirement $\phi_b=\pi-\phi_N$.
The next possible function is $\chi=\bar{b}+\frac{\bar{N}^{2}}{\bar{N}+\bar{b}\eta}$. The choice of $\eta$ that leads to the simplest solutions for the singularities is $\eta=1/4$. In this case the solution to $\chi=0$ still has a linear dependence on $N$. In this case $\chi=\infty$ is also a possible solution for $b\neq\infty$. All the equations for the solutions of the singularities for $\eta=1/4$ are also listed in Table~\ref{tab:poles}. Since the $\chi=0$ is also proportional to $-\bar{N}$ the same requirement $\phi_b=\pi-\phi_N$ applies. The solutions for the singularities can be seen in Fig.~\ref{fig:contour-chi}(b).
If the function is generalized to any value of $\eta$, the solution for $\chi=0$ is no longer linear. The angle $\phi_N$ can not be chosen freely and is restricted by $\pi/2<\phi_N<\pi-\arctan\left(\sqrt{4\eta-1}\right)$ \cite{Kulesza:2003wn}. The lower boundary is a usual restriction, however the upper limit is an additional restriction. The upper limit is a requirement in order to prevent the two lines for $\chi=0$ to intersect in positive real part of the plain as can be seen in Fig.~\ref{fig:contour-chi}(c). In order for the $b$ line to remain parallel to this line the angle needs to be chosen as $\phi_b=\pi-\arctan\left(\sqrt{4\eta-1}\right)-\phi_N$.

\begin{table}
	\begin{centering}
		\begin{tabular}{|c|c c|}
			\hline 
			$\chi$ & $\bar{b}+\bar{N}$ & $\bar{b}+\frac{\bar{N}^{2}}{\bar{N}+\bar{b}/4}$\tabularnewline
			\hline 
			$0$ & $\bar{b}=-\bar{N}$ & $\bar{b}=-2\bar{N}$\tabularnewline
			
			$\infty$ & \textbf{X} & $\bar{b}=-4\bar{N}$\tabularnewline
			
			$\rho_{L}$ & $\bar{b}=\rho_{L}-\bar{N}$ & $\bar{b}=\left(-4\bar{N}+\rho_{L}\pm\sqrt{\rho_{L}\left(8\bar{N}+\rho_{L}\right)}\right)/2$\tabularnewline
			\hline 
		\end{tabular}
		\par\end{centering}
	
	\caption{Different solutions of the equations for the singularities for some examples of the
		function $\chi$.}
	\label{tab:poles}
\end{table}

\phantomsection
\addcontentsline{toc}{section}{References}
\bibliographystyle{jhep}
\bibliography{biblio}
\end{document}